\documentclass[10pt]{aastex}
\usepackage{bm}
\usepackage{lscape}
\usepackage[dvips]{epsfig}
\usepackage{amsmath}
\usepackage{amssymb}
\usepackage{amsthm}
\usepackage{array}
\usepackage{multirow}

\begin{document}

\renewcommand{\baselinestretch}{1.0}
\newcommand{\ms}{$M_{\odot}$}
\newcommand{\msb}{$M_{\odot}$~}
\newcommand{\al}{$^{26}$Al}
\newcommand{\alb}{$^{26}$Al~}
\newcommand{\fe}{$^{60}$Fe}
\newcommand{\be}{$^{10}$Be}
\newcommand{\ca}{$^{41}$Ca}
\newcommand{\mn}{$^{53}$Mn}
\newcommand{\pd}{$^{107}$Pd}
\newcommand{\tc}{$^{99}$Tc}
\newcommand{\pu}{$^{244}$Pu}
\newcommand{\hf}{$^{182}$Hf}
\newcommand{\ct}{$^{13}$C}
\newcommand{\cm}{$^{247}$Cm}
\newcommand{\tho}{$^{230}$Th}
\newcommand{\nb}{$^{92}$Nb}
\newcommand{\msun}{\ensuremath{M_\odot}}
\newcommand{\mdot}{\ensuremath{\dot{M_6}}~}
\newcommand{\cd}{$^{12}$C}
\newcommand{\cdb}{$^{12}$C~}
\newcommand{\ctb}{$^{13}$C~}
\newcommand {\ket}[1] { |  #1 \rangle   }
\newcommand {\bra}[1] {  \langle #1  | }
\renewcommand{\thetable}{\arabic{table}}

\title{\large Theoretical Estimates of Stellar e$^-$ Captures.\\
I. The half-life of $^7$Be in Evolved Stars}
\author{S. Simonucci\altaffilmark{1}, S. Taioli \altaffilmark{2}, S. Palmerini \altaffilmark{3}, M. Busso\altaffilmark{4}}
\altaffiltext{1}{Department of Physics, University of Camerino, Italy}
\altaffiltext{1}{Istituto Nazionale di Fisica Nucleare, Sezione di Perugia, Italy;
stefano.simonucci@unicam.it}
\altaffiltext{2}{Interdisciplinary Laboratory for Computational Science,
FBK-CMM and University of Trento, Italy;
taioli@fbk.eu}
\altaffiltext{2}{Istituto Nazionale di Fisica Nucleare, Sezione di Perugia}
\altaffiltext{2}{Department of Physics, University of Trento, Italy}
\altaffiltext{2}{Department of Chemistry, University of Bologna, Italy}
\altaffiltext{3}{Departamento de F\'isica Te\'orica y del Cosmos, Universidad de Granada, Spain}
\altaffiltext{4}{Department of Physics, University of Perugia, Italy}
\altaffiltext{4}{Istituto Nazionale di Fisica Nucleare, Sezione di
Perugia, Italy}

\begin{abstract}
\begin{singlespace}
The enrichment of Li in the Universe is still unexplained, presenting various puzzles to
astrophysics. One open issue is that of obtaining reliable estimates for the rate of
e$^-$-captures on $^7$Be, for $T$ and $\rho$ conditions different from the solar ones. This
is of crucial importance to model the Galactic nucleosynthesis of Li. In this framework, we
present here a new theoretical method for calculating the e$^-$-capture rate in conditions
typical of evolved stars. Furthermore, we show how our approach compares with state-of-the-art
techniques for solar conditions, where various estimates are available. Our computations include: i)
``traditional'' calculations of the electronic density at the nucleus, to which
the e$^-$-capture rate for $^7$Be is proportional, for different theoretical approaches
including the Thomas--Fermi, Poisson--Boltzmann and Debye--H\"uckel (DH) models of screening,
ii) a new computation, based on a formalism that goes beyond the previous ones, adopting
a mean-field ``adiabatic'' approximation to the scattering process. The results obtained
with the new approach as well as with the traditional ones and their differences are
discussed in some detail, starting from solar conditions, where our approach and the DH model
essentially converge to the same solution. We then analyze the applicability
of both our method and the DH model to a rather broad range of $T$ and $\rho$ values,
embracing those typical of red giant stars, where both bound and continuum states contribute
to the capture. We find that, over a wide region of the parameter space explored, the DH approximation does not really stand, so that the more general method we suggest should be preferred. As a first application, we briefly reanalyze the $^7$Li abundances in RGB and AGB stars of the Galactic Disk in the
light of a revision in the Be-decay only; we however underline that the changes we find in
the electron density at the nucleus would induce effects also on the electron screening (for
p-captures on Li itself, as well as for other nuclei) so that our new approach might have
rather wide astrophysical consequences.
\end{singlespace}
\end{abstract}
\keywords{Poisson--Boltzmann, Debye--H\"uckel, Thomas--Fermi models - first-principles methods - Stars: RGB, AGB - abundances - Nuclear reactions - Nucleosynthesis - Electron screening}

\section{Introduction}
Serious problems affect our understanding of the Li evolution in the Galaxy.
Big Bang Nucleosynthesis \citep[BBN: see e.g. ][and references therein]{Coc}
predicts a Li abundance  higher than observed in extremely metal-poor objects
\citep{Bonifacio} and in low-metallicity Main Sequence (MS) stars \citep{Spite}.
On the other hand, in the present interstellar medium (ISM)
the $^7$Li abundance is higher even than that expected by BBN \citep{Casuso}.
As Galactic Cosmic Rays do not produce much $^7$Li \citep{fields,alibes} we should rely on
stellar nucleosynthesis for explaining this increase.

$^7$Li is however a very fragile nucleus, so that its photospheric concentration
in stars is easily destroyed when convective processes can carry it to
moderately high temperatures (a few millions K), where it undergoes
proton captures \citep{Boesgaard1}. Special conditions are therefore required for Li production,
which are limited to a few astrophysical scenarios. These mainly include novae and
intermediate mass stars (IMS) undergoing H burning at the base of their envelope through the so-called Hot Bottom Burning process \citep[see e.g.][]{dv10}.
In current stellar models for low mass stars (LMS: those with masses below about $2 - 3$ $M_{\odot}$)
Li is instead predicted to be destroyed already in the early phases of
evolution, preceding the MS \citep{Pinsonneault,Sestito}. In stars of mass lower than
solar, whose convective envelopes remain large even during the MS,
Li destruction continues through this stage; for higher masses, instead,
it is predicted that external convection shrinks and does not
include any more zones hot enough to affect Li. Contrary to these
model expectations, observations of the Sun and of solar-like stars
reveal that they undergo extensive Li-depleting processes during
central hydrogen burning. One of the consequences is that the solar
photosphere is about 100 times less Li-rich than meteorites \citep{Asplund}. Li-destruction processes are known to occur also at intermediate
effective temperatures, in MS stars of the Galactic disc,
generating the so-called Li{\it -dip} \citep[see e.g.][]{Boesgaard2,Balachandran,Boesgaard3}.

The above phenomena, unexpected from canonical stellar models,
have been interpreted (sometimes only qualitatively) in terms of mixing episodes of a nature
different from pure convection \citep{Michaud1,Michaud2},
often attributed to rotationally-induced effects \citep{Charbonnel1}, atomic diffusion,
or magnetic dynamo processes \citep{Eggenberger}, like those observed in active stars
\citep{andrews}. However, the very large Li destruction in the Sun and in other main sequence stars is still not accounted for in detail, being too small in the models.
In more advanced stages, i.e. along the Red Giant Branch
(hereafter RGB) and along the Asymptotic Giant Branch (hereafter
AGB) similar phenomena must be active \citep{Palmerini1,maiorca},
perhaps again related to magnetic effects \citep{Busso,Nordhaus,Denissenkov}.
The consequence is that, in most evolved stars, Li is further depleted as compared to
MS stages \citep[see e.g.][and references quoted
therein]{Palmerini2}.

At odds with our need for finding sites where Li is produced in stars,
only few red giants ($\sim 2\%$) show Li enhancement at their photosphere
 \citep[see e.g.][]{Brown1,Charbonnel2,Kumar,Uttenthaler,lebzelter}.
Their abundances might in principle be produced by
coupled mixing and nucleosynthesis episodes, in which the depletion of Li by
downward diffusion is over-compensated by an upward transport of $^7$Be from burning regions,
at a fast enough pace that it survives destruction by $p$ and $e^-$
captures, reaches the stellar envelope and finally decays there, reproducing Li.
The situation is however far from being clear \citep{Charbonnel1,Palmerini2}.
Quantitative modeling is in particular hampered by a poor knowledge of how
the rate of Be decay changes in the rapidly varying conditions below the envelopes of red giants.
In fact, extrapolations of this rate from the works done for the Sun \citep{bahcall1962,Iben,bm69} are extremely insecure, due the ambient conditions of H burning in evolved stars, which
are very different from solar. Indeed, in the layers above the H-burning shell of a red giant
the temperature ($T$) spans a range from 70-80 MK down to a few MK, while the density ($\rho$)
is lower than in the solar center from one to five orders of magnitudes.

In general, the pioneering works of the sixties for the Sun were performed considering
the ionization degree of Be through the Saha equation, and including the contribution of
free electrons on the assumption that inside a Debye radius around a Be nucleus
they behave as a Maxwellian gas, thus following a treatment of the Coulomb screening
in the plasma originally due to Debye and H\"uckel \citep{Debye}.
Although more recent and general approaches do exist
\citep{gruzinov,Brown2,Sawyer} sometimes demonstrating
the limits of the Debye--H\"uckel (hereafter DH) approximation \citep{Johnson}, they maintain a
treatment similar to the Born-Oppenhemier approach, where the electronic response is considered
to be much faster than the ionic one. These works almost invariably find for the Sun results very similar to those of the classical studies cited above, but they do not consider the physical conditions prevailing in evolved stars.

It has to be further noted that, at the high temperatures of shell H-burning in
red giants, recombination might occur in states that are highly excited and very close to each other.
In such conditions, the Born-Oppenheimer approach is questionable; moreover, the conditions for
the classical DH approximation often do not actually hold (see next sections) and one does not know whether this introduces small or large deviations in the capture rate.

We decided therefore to explore the problem of $e^-$ captures on $^7$Be for the typical
$T$ and $\rho$ values of H-burning layers in evolved stars, both following the traditional
DH approach and introducing a new treatment, in which
the assumption that electrons follow a Maxwell-Boltzmann energy distribution is relaxed
(considering them, more generally, as a Fermi gas) and in which a mean-field adiabatic
approximation to the scattering process is used. The primary scope is to provide the missing
weak-interaction input data for Li nucleosynthesis calculations, clarifying whether
such data can be deduced in the traditional way or not. Subsequently, we also plan to apply our
results to a re-evaluation of the problem of electron screening in stellar plasmas (see some
comments on that issue in sections 3.1 and 5).

We remind that other cases exist, involving nuclei heavier than Li,
in which the predictions of nucleosynthesis are still unreliable for the
lack of knowledge about the dependence of electron captures on the ambient conditions,
during the transport of newly produced nuclei to the stellar surface. Crucial
isotopes subject to these uncertainties are, e.g., $^{41}$Ca and $^{205}$Pb, which are important clocks for dating the latest nucleosynthesis
processes before the contraction of the Solar Nebula \citep[see][and references therein]{Busso2, Busso3}. If one considers
also $\beta$ decays, then it turns out that several reaction branchings along the $s$-process
path are still affected by our poor knowledge of weak interactions in stars, in contrast
with the high accuracy of the competing neutron-capture processes.

In general, a better knowledge of radioactive decays would be
relevant for a large number of physical problems, well beyond the
borders of stellar astrophysics. Accurate decay rates are e.g. of paramount
importance in Earth and Planetary sciences, in order to date
geological and astronomical processes by estimating the amount of a
given long-lived species remaining in a sample (a rock or a stellar
photosphere). Furthermore, nuclear decay provides an impressive source of heat
in any planetary body, including the Earth. It is believed that as
much as half the heat measured at the Earth surface, corresponding
to approximately 21 TW, be due to radioactive processes involving
$^{40}$K, $^{232}$Th, $^{235}$U and $^{238}$U, occurring both in the
crust and in the core \citep{Pollack,Fowler,Lee}. Heating
from shorter-lived radioactivities like $^{26}$Al
is then believed to provide the energy for melting and
differentiating the early solid bodies around the Sun \citep[see e.g.][and
references therein]{Busso3}. In a forthcoming paper our results will therefore
be extended to other nuclei and applied to the clarification of a few such
problems.

This paper is structured in the following way.
In section 2 we provide a general introduction to the current problems
of the electron captures on $^7$Be.
In section 3 we discuss the theoretical framework used for solving the ($^7$Be + $e^-$)
scattering event, referring to Appendix A for a more detailed technical discussion.
In the same section 3 we present calculations of the electron density at
the $^7$Be nucleus, obtained by applying several models available in the literature
(as detailed in Appendix B) and we show comparisons among such values. Section 4
then illustrates in a preliminary way some consequences for the problem
of Li production and destruction in evolved stars, while tentative
conclusions are drawn in section 5.

\section{The electron captures on $^7$Be: state-of-the-art.}

The driving force responsible for the electron capture decay is the weak nuclear
interaction, a process of very short range that was therefore long believed
to be insensitive to extra-nuclear factors, notably the chemical environment,
the ionization degree, the pressure and temperature conditions. Contrary to this simple
view, many authors reported evidence of changes in nuclear decay
rates with temperature \citep{emery1972,Hahn}, pressure
\citep{Lee}, and the chemical environment \citep{ray2002,Ohtsuki, Morisato},
which are believed to be connected to the modification of the electron
density at the nucleus, $\rho_e(0)$, induced by the change of these parameters.
The rate indeed depends on the s-type atomic-orbital wavefunctions, the only ones with finite
density at the nucleus. Therefore, factors affecting this density can appreciably modify
the decay rate. Besides $T$ and $\rho$ values, also the pressure, the level of ionization,
and the presence of other charged particles, screening the interaction, might in principle
be of relevance. On the other side, ab-initio calculations of the radioactive decay
rate at room temperature for $^7$Be (and several other isotopes) performed by \citet{Lee}
showed a very small dependence (within $\sim 0.1 - 0.2$\%) on the chemical environment
and on the pressure, up to 25 GPa.\\
\indent Furthermore, very little is still known about decay events that
occur at very high $T$ in ionized media for changing $\rho$, such as
those found in stellar interiors, and similar studies are still challenging for theory
and intensively debated.
The recent recommendations by \citet{Adelberger} are mainly based on
the work by \citet{bm69} and \citet{Iben}.
In these works a partial ionization of $^7$Be in the Sun was assumed, thus
the rate now currently used includes contributions from both bound and continuum states
and the total-to-continuum capture ratio is 1.217.
Different results were presented by \citet{Shaviv}, who
assumed that $^7$Be is fully ionized in the solar plasma; this
implied an increase of the $^7$Be lifetime by $\sim$ 20 to 30\% as
compared to previous recommendations. An even more complicated
situation is depicted by \citet{Quarati}, who recently
found a $^7$Be lifetime shorter by about 10\%, using a modified
DH screening potential. The present situation is therefore
quite unsatisfactory and the uncertainties affecting Li abundances in stars have
to cope also with this poor understanding of the basic
nuclear input data. \\

\indent This is actually the main motivation of our present
attempt, as essentially no one of the existing reaction rate compilations
for stellar physics can be safely extrapolated from the Sun to other situations.
For example, the rate by \citet{Adelberger} is derived from a simple fit
over a very small domain of the parameter space, while the formula by \citet{Fowler1975}
imposes a specific choice for the density.

In this context, our goal is to lay the foundations of a theoretical
and computational method for studying the electron capture and decay
rates at high temperature in a density-varying medium, to go beyond
the existing treatments. We shall probe its validity through a comparison
with more classical approaches, and shall use the results to reconsider briefly
the problem of Li production and destruction in red giant stars.

\section{Dynamics of electron capture from {\it ab-initio} calculations: our model}

In this section we describe our first-principle approach for computing electron captures
over a much wider range of $T$ and $\rho$ than so far
possible. These two parameters alter considerably the balance between bound and
free electronic states contributing to the capture.\\
\indent While we believe that our formalism is totally general and can be applied even to other systems,
as a test case of our method we will estimate the decay rate of $^7$Be by electron captures:
\begin{equation}\label{reaction}
^{7}{\text Be} + e^{-} \rightarrow ~ ^{7}{ \text Li} + \nu_e
\end{equation}
$^{7}{\text Be}$ can decay into $^{7}{ \text Li}$ through different decay
channels. In particular, the decay from the ground state of $^{7}{\text Be}$ can
occur to both the ground and the first excited state of $^{7}{ \text Li}$. Of
course the phase space and the kinetic energy of the neutrinos will be different
for these two decay paths. At ambient conditions, i.e. for negligible
kinetic energy of the impinging electron, $^{7}{\text Be}$ decays in 53 days into
the ground state of $^{7}{ \text Li}$ (3/2$^-$) for the 89.7\% of cases, while for the
remaining 10.3\% it decays into the first excited state (1/2$^-$), as discussed by  \citet{Mathews}.
The energy of this latter is 477.4 KeV higher than for the ground state. The weak branching ratio (BR) at room temperature is therefore 8.709. By defining Q$_0$ and Q$_1$ as the kinetic energies of the neutrinos, escaping respectively from the $^{7}$Li nucleus in its ground and in its first
excited state, we have Q$_0$ = 861.6 keV and Q$_1$ = Q$_0$ - 477.4 = 384.2 keV \citep{Fuller}. Being the kinetic energy higher in the first case, the available phase space will be larger. Roughly we can estimate that, at a temperature $T = 10^7$ K, BR$=89.7/10.3\times(Q_0+kT)^2/(Q_1+kT)^2/(Q_0^2/Q_1^2)=8.684$. The percentage variation of the BR due to an increase of the temperature by five orders of magnitude is thus only 0.3\%. Of course, even the weak matrix elements will be different for these two channels and this should be in principle taken into account. However, our main goal here is to estimate accurately the total electron capture decay rate, as this is what we need for assessing the astrophysical consequences of our new results. Thus, in the following discussion we shall assume that the decay occurs only to the ground state of $^{7}{\text Li}$.

The framework within which we shall calculate the decay rate is given by the
theory of scattering under two potentials: $V$, representing the screened,
short-range Coulomb potential and $W$, which represents the weak interaction
coupling the Coulomb distorted initial state and the final channels.
We define $\phi_{i,\bm{p}}$ as a free plane-wave, $\phi_{i,\bm{k}}^{+}$ the
perturbed `in-state', described by a Coulomb-distorted plus an outgoing
spherical wave, $\phi_{f,\bm{k}}^{-}$ and $\psi_{f,\bm{k}}^{-}$ the perturbed
`out-states', which describe asymptotically the emission of a neutrino with
relative momentum $\bm{k}$, released into the final channel $f$ of the target,
respectively with and without the weak coupling. Then we can write the cross section
of the electron capture process as:
\begin{eqnarray}\label{cs}
\sigma_{i\rightarrow f} & = & \int\frac{d^{3}k}{(2\pi)^{3}}\frac{2\pi}{v}\left|\left\langle \psi_{f,\bm{k}}^{-}|W|\phi_{i,\bm{p}}^{+}\right\rangle +\left\langle \phi_{f,\bm{k}}^{-}|V|\phi_{i,\bm{p}}\right\rangle \right|^{2}\delta\left(\frac{p^{2}}{2m_e}+E_{i}-E_{f}-ck\right)\nonumber \\
 &  & =\int\frac{d^{3}k}{(2\pi)^{3}}\frac{2\pi}{v}\left|\left\langle \phi_{f,\bm{k}}^{-}|T_{w}|\phi_{i,\bm{p}}^{+}\right\rangle \right|^{2}\delta\left(\frac{p^{2}}{2m_e}+E_{i}-E_{f}-ck\right)
\end{eqnarray}
$E_i, E_f$ represent the internal energies of the
target ($^{7}$Be) and of the final decay product ($^{7}$Li),
$\bm{p}=m_ev$ and $\bm{k}$ are the relative electron and neutrino
momenta in the initial and final channels, and $v$ is
the electron velocity in the initial channel, relative to $^{7}$Be.
In Eq. (\ref{cs}) the matrix element
$\left\langle\phi_{f,\bm{k}}^{-}|V|\phi_{i,\bm{p}}\right\rangle$ must vanish, as
the Coulomb interaction does not couple the initial and final
decay channels ($\left\langle\phi_{f,\bm{k}}^{-}|V|\phi_{i,\bm{p}}\right\rangle = 0$). Thus we can define the $T$-matrix of the weak interaction as
\begin{equation}\label{elmat}
\left\langle
\psi_{f,\bm{k}}^{-}|W|\phi_{i,\bm{k}}^{+}\right\rangle
=\left\langle
\phi_{f,\bm{k}}^{-}|{\text{T}_{W}}|\phi_{i,\bm{k}}^{+}\right\rangle
\end{equation}
From Eq. (\ref{cs}) one can obtain the electron capture rate by multiplying the
cross section for the electron current:
\begin{equation}\label{rate}
\Gamma_{i\rightarrow f}=\int2\pi\frac{d^{3}k}{(2\pi)^{3}}\left|\left\langle \phi_{f,\bm{k}}^{-}|T_{w}|\phi_{i,\bm{p}}^{+}\right\rangle \right|^{2}\delta\left(\frac{p^{2}}{2m_e}+E_{i}-E_{f}-ck\right)
\end{equation}
The calculation of the nuclear $T$-matrix elements, particularly with the inclusion of the
$p^{2}/{2m_e}$ dependence, is very difficult from first-principles and is out
of the scope of this paper.  Indeed $p^{2}/{2m_e}$ depends on the temperature
of the system.
However, the full calculation is not needed in our case, and two approximations will enable us
to simplify the equations without sacrificing accuracy. Firstly, we will neglect the dependence
on the temperature of this matrix element, and consider only one initial state.
This approximation is motivated by the fact that the first nuclear excited
state of $^7$Be is at 429.2 keV above the ground state \citep{Fuller}, corresponding to a temperature of 5$\times 10^9$ K, larger by factors of 50 to 100 with respect to the maximum values found during H-shell burning in the RGB or AGB phases.

The second approximation is to model the weak interaction, owing to its very
short-range nature, by a Fermi contact interaction $\text{T}_{W} \propto
\delta(\bm{r})$, independent of the neutrino momentum $\bm{k}$.
By integrating out in $\bm{k}$ using the latter assumption:
\begin{eqnarray}\label{gamma} \Gamma_{i\rightarrow f} & = &
\frac{\bar{k}^{2}}{\pi
c}\left|t_{f,i}\langle i,0|\phi_{i,\bm{p}}^{+}(0)\rangle\right|^{2}=\frac{1}{\pi
c^{3}}\left|t_{f,i}\right|^{2}\left\langle i,0|\phi_{i,\bm{p}}^{+}\right\rangle
\left(\frac{p^{2}}{2m_e}+E_{i}-E_{f}\right)^{2}\left\langle
\phi_{i,\bm{p}}^{+}|i,0\right\rangle \nonumber \\
 & = & \frac{1}{\pi c^{3}}\left|t_{f,i}\right|^{2}\left\langle
i,0|\phi_{i,\bm{p}}^{+}\right\rangle
\left(H_{0}+V+E_{i}-E_{f}\right)^{2}\left\langle
\phi_{i,\bm{p}}^{+}|i,0\right\rangle \end{eqnarray}
where $\bar{k}=\frac{1}{c}\left(\frac{p^{2}}{2m_e}+E_{i}-E_{f}\right)$ and $\langle\phi_{i,\bm{p}}^{+}|i,0\rangle$
is the electron wavefunction representation at the Be nucleus.
We will assume that the nuclear $T$-matrix elements $t_{f,i}$ are known and equal to those
measured on the Earth. The assumptions of Eq. (\ref{gamma}) imply that the electron capture can be
modeled as a $^7$Be-$e^-$  two-body scattering process at a given relative electron momentum $p$, and
that the rate is proportional to the electron density at the nucleus $\rho_e(0)$, which is screened and modified
by the presence of the surrounding particles.\\
\indent Our model system of the hot plasma, for different conditions of $T$ and $\rho$,
is represented by a homogeneous Fermi gas composed by $^7$Be atoms surrounded
by $N_p$ protons (hydrogen nuclei) and $N_e$ electrons. The presence of other particles,
such as helium, will be neglected in the following discussion and in the
stellar decay rate calculation, unless otherwise stated.
However, the generalization of our method to include several species is straightforward. \\
\indent An {\it exact, ab-initio} calculation of the electron capture
rate would be extremely complex, due to the many-body nature of the scattering, and
one needs to introduce some further approximations.
The first step of our method is thus the formal reduction
of this complicated many-body problem to a screened two-body scattering problem, by using
an ``adiabatic'' factorization of the eigenfunctions, resembling the widely known Born-Oppenheimer (BO)
approximation adopted in standard electronic structure methods.
This goal is reached by fixing the reference frame into the $^7$Be nucleus and
by writing the different parts of the Hamiltonian in this non-inertial frame (we remember that the
$^7$Be nucleus is in principle free to move around).
In every sense this is only a coordinate transformation. However, as in
classical mechanics, the consequence of using a non-inertial frame will bring
about some complicacy (apparent forces), e.g. in this case
we will have a two-body kinetic energy operator.
In the appendix we describe the mathematical details of this derivation, from which we obtain
that the many-body scattering Hamiltonian in the coordinate system relative to the $^7$Be nucleus
can be written as:
\begin{eqnarray}\label{Hamiltonian}
&& H = \sum_{j=1}^{N_{e}}\left(-\frac{1}{2m_{e}}-\frac{1}{2M_{Be}}\right)\nabla_{e,j}^{\prime2}+\sum_{J=1}^{N_{p}}\left(-\frac{1}{2m_{p}}-\frac{1}{2M_{Be}}\right)\nabla_{p,J}^{\prime2}-\sum_{j=1}^{N_{e}}\frac{Z_{Be}}{|\bm{r}_{e,j}^{\prime}|}+\sum_{J=1}^{N_{p}}\frac{Z_{Be}}{|\bm{R}_{p,J}^{\prime}|}\nonumber \\
 && - \sum_{j=1}^{N_{e}}\sum_{J=1}^{N_{p}}\frac{1}{|\bm{r}_{e,j}^{\prime}-\bm{R}_{p,J}^{\prime}|}+\sum_{j=1}^{N_{e}}\sum_{k=j+1}^{N_{e}}\frac{1}{|\bm{r}_{e,j}^{\prime}-\bm{r}_{e,k}^{\prime}|}+\sum_{J=1}^{N_{p}}\sum_{K=J+1}^{N_{p}}\frac{1}{|\bm{R}_{p,J}^{\prime}-\bm{R}_{p,K}^{\prime}|} - \frac{1}{2M_{Be}}\nabla_{Be}^{\prime2} \nonumber \\
&& - \sum_{\substack{J,J^{\prime}=1 \\J \neq J^{\prime}}}
^{N_{p}}\left( \frac{1}{M_{Be}}\bm{\nabla}_{p,J}^{\prime}\cdot\bm{\nabla}_{p,J^{\prime}}^{\prime}\right) -\sum_{\substack{j,j^{\prime}=1 \\j \neq j^{\prime}}}^{N_{e}}\left(\frac{1}{M_{Be}}\bm{\nabla}_{e,j}^{\prime} \cdot \bm{\nabla}_{e,j^{\prime}}^{\prime}\right)-\frac{1}{M_{Be}}\sum_{j=1}^{N_{e}}\sum_{J=1}^{N_{p}}\bm{\nabla}_{p,J}^{\prime}\cdot\bm{\nabla}_{e,j}^{\prime} \nonumber \\
&& + \sum_{j=1}^{N_{e}}\left(\frac{1}{M_{Be}}\bm{\nabla}_{e,j}^{\prime}\cdot\bm{\nabla}_{Be}^{\prime}\right)+\sum_{J=1}^{N_{p}}\left(\frac{1}{M_{Be}}\bm{\nabla}_{p,J}^{\prime}\cdot\bm{\nabla}_{Be}^{\prime}\right)\nonumber \\
&&
\end{eqnarray}
where, in order of appearance, one has the electron and proton kinetic
energies, the electron-Be and proton-Be potential energies,
the electron-proton, electron-electron, proton-proton
interaction, the Be kinetic energy and, finally, terms
coupling the different particle species.
In the inter-particle coupling terms, resulting from our
coordinate transformation, $\left\{ \bm{r}_{e,j}^{\prime},
\bm{R}_{p,J}^{\prime}\right\}$ identify the coordinates of the $j$-electron and
$J$-proton relative to the $^7$Be coordinate, $\bm{R}_{{\text
Be}}=\bm{R}_{\text{Be}}^{\prime}$.

We then look for separable eigensolutions of the form:
\begin{eqnarray}\label{factor}
\Psi\left(\bm{R}_{{\text Be}}^{\prime},\left\{ \bm{r}_{e}^{\prime}\right\} ,\left\{ \bm{R}_{p}^{\prime}\right\} \right)&=&\chi(\bm{R}_{{\text Be}}^{\prime})\Phi\left(\left\{ \bm{r}_{e}^{\prime}\right\} ,\left\{ \bm{R}_{p}^{\prime}\right\} \right)\\
\bm{\nabla}_{\text Be}^{\prime}\chi(\bm{R}_{\text Be}^{\prime})&=&\bm{k}\chi(\bm{R}_{\text{Be}}^{\prime})
\end{eqnarray}
This wavefunction factorization differs from the usual formulation of the BO approximation
in two ways. At variance with the BO approximation, the function
$\Phi\left(\left\{\bm{r}_{e}^{\prime}\right\} ,\left\{
\bm{R}_{p}^{\prime}\right\} \right)$
in Eq. (\ref{factor}), written in the $^7$Be reference system, does not depend parametrically
on the $^7$Be coordinates and thus needs to be calculated only once. Furthermore, the BO
approximation is applicable only when the electronic potential energy surfaces are well
separated. These are not our conditions,
as the electrons, for $T$ and $\rho$ values pertinent to the burning
regions of stars, occupy either highly excited states of the $^7$Be atom or
continuum orbits. Thus, we are out of
reach of the BO scheme, which cannot be rigorously applied.

However, it can be shown that the last two coupling terms in Eq. (\ref{Hamiltonian}), which are
crucial for the application of our mean-field treatment of the many-body interaction, can
be neglected. In Appendix A we provide a detailed explanation of the conditions where this
approximation can be used.

We notice that, by introducing the two-body framework and the relative coordinate
system, we can achieve two important results. The first one is that in our approach
the $^7$Be nucleus is in principle free to move around (thus overcoming the limitations of
the Born-Oppenheimer scheme), even though in general its motion is strongly limited by the
presence of other particles; the second one is that, by neglecting the two last terms of Eq.
(\ref{Hamiltonian}), the two-body electron-$^7$Be density-matrix can be factorized as the
product of two one-body density-matrices and thus it is possible to introduce different
schemes of approximations to the many-body interaction, including the Hartree--Fock's one
(hereafter HF). Therefore, the screening brought about by all the interacting fermions of
the surrounding environment, which modifies the two-body electron-$^7$Be scattering and
thus the electron-capture rate, can be now taken into account, using standard many-body
techniques.

\subsection{Screening at different levels of accuracy: Thomas--Fermi, Poisson--Boltzmann and Debye--H\"uckel}

While the importance of the screening for the assessment of the electron-capture rate is well understood,
the approaches used so far are all based on the DH approximation;
thus, their reliability is not a priori guaranteed in every situation.
Within this approximation \citet{Iben}, for example, realized that, in solar conditions,
the electron density at the nucleus is reduced by the electronic screening
for both bound and continuum electrons.
\citet{gruzinov} further improved this model by integrating the
density-matrix equation to treat on similar grounds the bound and continuum electrons, and
by including via a Monte-Carlo approach non-spherical charge fluctuations induced
by the small number of ions within the radius $\lambda_{D}$ of a Debye sphere.\\
\indent Despite these improvements, the conclusions drawn by these works
rely on the assumption that the hot plasma can be modeled as a classical
non-interacting electron gas, screened by using the DH model,
in thermodynamic equilibrium with a heat bath at absolute temperature $T$.
The range of applicability of this approximation is based on both classical and
statistical considerations: for the former ones, we need a large number of electrons and a high
temperature;  for the latter ones, we require a smooth change of the potential
over a characteristic distance ($\lambda_{D}$), which is large as compared to
the thermal De-Broglie wave-length of the electrons ($\lambda_{DB}$).  However,
in the solar case, where $T \simeq 16 \times 10^6 K, \rho = 150 g/cm^{3}$,
we have, for the electrons, $\lambda_{D} = 0.407, \lambda_{DB}=0.352$ a.u.. The
conditions are therefore, already for our Sun, at the limits of validity of the
classical (Maxwell-Boltzmann) gas approximation. Over the more extended range of
parameters characterizing the radiative layers above the H-shell in a red giant
(down to $T\simeq 2.0 \times 10^6 K$) the application of the Maxwell--Boltzmann
distribution might no longer be justified.

We underline that the simple DH approach, commonly used
in the literature so far, can actually be derived as a two-step
approximation of the more general Thomas--Fermi (hereafter TF) model, by using the linearized Maxwell-Boltzmann distribution. (In Appendix B we provide a detailed discussion of the DH treatment).

\indent Hence we can compute the crucial parameter, i.e. the electron density at the Be nucleus $\rho_e(0)$, at different levels of approximation, in order to disentangle the differences introduced in the results by the various approaches.  In Table 1 we report the results of such calculations, adopting as an example the solar conditions. We underline that $\rho_e(0)$ is directly related to the electron-capture rate.

In the above calculations, the plasma was modelled by a $^7$Be nucleus surrounded by a Fermi gas occupying an infinite volume and helium was explicitly included.
One can see that both the classical Poisson--Boltzmann approach (PB) and the DH
approximation give sizeable different results (of the order of 38\% and 6\%
respectively) with respect to the TF approximation, already for solar conditions. The change becomes even larger when we use our more general method, based on the HF approximation, which we will discuss in the next section. (Surprisingly, the DH approximation is closer to the HF and TF results than the PB case, from which it is derived by linearization). \\

\begin{table}[hbt!]
\centering

{\large Table 1}

\begin{tabular}{|c|c|c|c|}
\hline
$\rho_{HF}(0)$ & $\rho_{TF}(0)$ & $\rho_{PB}(0)$ & $\rho_{DH}(0)$\tabularnewline
\hline
$16.320 \pm 0.020$ & $16.135 \pm 0.025$ & $14.870 \pm 0.020$ & $16.085 \pm 0.025 $\tabularnewline
\hline
\end{tabular}
\caption{Values (in atomic units) of the electron density at the Be nucleus $\rho_e(0)$ for different levels of the theory (HF=Hartree--Fock, TF=Thomas--Fermi, PB=Poisson--Boltzmann, DH=Debye--H{\"u}ckel) for solar conditions ($T=16$MK).
The conditions at which we performed the calculations are $T = 15.67$MK, $\rho$=152.9 g/cm$^3$,
$X_{\rm H}$ = 0.34608, $X_{\rm He}$ = 0.63368.}
\end{table}

\indent Even more important, one must underline that the PB and DH approximations
do not hold outside the conditions of the solar nucleus. This is true, in particular, at lower temperatures and densities, where a large part of the Li production occurs (because the competing p-captures on $^7$Be become ineffective). In order to check this point, we performed the calculations of the the electron density at the Be nucleus for all the previously discussed approaches over a wide range of $T$ and $\rho$ conditions. The results are reported in Table 2, where $T$ is in units of 10$^6$K and $\rho$  in $g/cm^{3}$. Values at the left of the vertical bar ($|$) represent the electron density at the nucleus
$\rho_e(0)$, while those at the right represent $\rho_e(0)$ multiplied by
$(1 + p^2/2m)/(E_i-E_f)$ (see Eq. \ref{gamma}), to which
the capture rate is proportional.  The Debye radius $\lambda_D$ and the
De-Broglie wavelength $\lambda_{DB}$ for electrons ($e^-~$) and protons ($p$)
are given in atomic units.

From the analysis of the electron density at the nucleus $\rho_e(0)$ for the
different approximations, it is clear that, when we are out of reach of the DH
approximation ($\lambda_D \le \lambda_{DB}$), the values calculated using this
model may differ by more than 30\% from the HF values. In particular, we
underline that, except for very high temperatures, the PB approximation is
extremely unreliable, while over the whole considered range of $T$ and $\rho$
values, the TF model systematically underestimates the density at the nucleus
$\rho_e(0)$. At variance with this, the DH model, except for low temperatures
($T\le 10^6$ K), seems to partially correct the wrong behavior of the PB
approximation, but not in a systematic way. This is due to the fact that the
electron density at the nucleus $\rho_e(0)$ is finite for the DH model, while
it is infinite for the Boltzmann approximation (in the HF method the density is
always finite, due to the fact that in this case the electronic wavefunction is
finite at the nucleus).
It is thus evident that the DH model fails in capturing the electron screening
properties over a portion of the parameter space that is very important for Li
nucleosynthesis and, if one is to assess accurately the electron-capture rate
of $^7$Be in that region, it is necessary to go beyond this approach.

We notice that the issue at stake here is not only a better estimate of
weak interactions. If the traditional methods underestimate the electron
density over a considerable part of the conditions typical of H-burning in Red Giants, as it seems to be the case, then this implies that they also underestimate the effect of electrons in screening strong interactions. In the case we discuss here (Li production and destruction) our next step will therefore be a re-evaluation of electron screening for proton captures on $^7$Li, which occur down to very low temperatures (few 10$^6$K). It is well known that the huge destruction of Li in the Sun is still unaccounted for quantitatively, being too small in the models, so that a more effective process of proton captures on Li, induced by an
increased estimate of the electron density and of its screening of the Coulomb
barrier, might have important consequences on solar physics and on our
understanding of nucleosynthesis in general.

At high temperature and high density ($T \ge 10 ^7$ K, $\rho \ge 100$ g/cm$^3$)
the approaches discussed so far (with the exception of the purely classic PB
treatment) become essentially equivalent to one another, as shown in Table 1;
in particular, the values of the electron density at the Be nucleus produced by
our new method and the one from the DH approximation differ by less than 1.5\% (see Table 1, column 1 and 4).

\subsection{Calculations of an accurate decay rate from first principles}

In order to apply our model Hamiltonian, Eq. (\ref{Hamiltonian}),
to the calculation of the capture rate over a wider range of
$T$ and $\rho$ values than so far possible, we need to develop new tools, suitable to go
beyond the DH approximation and capable of
treating accurately the electronic screening for both bound and continuum electrons.
We will then use the new estimates for interpreting
the observational data.

In our approach, the Coulomb screening is calculated using a
temperature-dependent HF method, within the canonical ensemble, by populating
both the ground-state and excited-state atomic orbitals via a Fermi-Dirac
distribution.
In our treatment, we will neglect the electron-proton pair-formation at
low temperature ($T < 10^6$ K). Our method can be outlined as
follows. i) We assume that the decay rate is proportional to $\rho_e(0)$
calculated by solving self-consistently a system of coupled HF equations for
protons and electrons in the electrostatic field of a $^7$Be nucleus at the origin of the
coordinate system. ii) The coupling is given by the self-consistent
Hartree term, describing the Coulomb interaction between the two fermionic
species. iii) The chemical potentials of protons and electrons are chosen in
such a way that the system is neutral far from the $^7$Be atom. iv) Within this
framework, the scattering of $^7$Be with protons and electrons is a two-body
interaction, with the remaining particles acting as a mean-field. \\
The first step in the above list consists in solving self-consistently the equations
for the electron charge density $\rho$ and the potential $V$ to calculate the electron-capture rate.
The HF equations for this problem read:
\begin{eqnarray}\label{hartree1}
-\frac{1}{2m_{j}}\nabla^{2}\psi_{j,\alpha\sigma_{e}\tau_{Be}n}(\bm{r})+V_{j}^{ext}(\bm{r})\psi_{j,\alpha\sigma_{e}\tau_{Be}n}(\bm{r})-\mu_{j}\psi_{j,\alpha\sigma_{e}\tau_{Be}n}(\bm{r})\label{eq:HF_eq_Be7}\\
+\sum_{\beta\sigma_{e}^{\prime}\tau_{Be}^{\prime}}\int
d\bm{r}^{\prime}V_{j,\alpha\sigma_{e}\tau_{Be}\beta\sigma_{e}^{\prime}\tau_{Be}^{\prime}}^{HF}(\bm{r},\bm{r}^{\prime})\psi_{j,\beta\sigma_{e}^{\prime}\tau_{Be}^{\prime}n}(\bm{r}^{\prime})
& = & \epsilon_{j,n}\psi_{j,\alpha\sigma_{e}\tau_{Be}n}(\bm{r})\nonumber
\end{eqnarray}

\begin{eqnarray}\label{hartree2}
\rho_{j,\alpha\sigma_{e}\tau_{Be}\alpha^{\prime}\sigma_{e}^{\prime}\tau_{Be}^{\prime}}(\bm{r},\bm{r}^{\prime}) & = & \sum_{n}\frac{1}{e^{\frac{\epsilon_{j,n}}{kT}}+1}\psi_{j,\alpha\sigma_{e}\tau_{Be}n}(\bm{r})\psi_{j,\alpha^{\prime}\sigma_{e}^{\prime}\tau_{Be}^{\prime}n}(\bm{r}^{\prime})
\end{eqnarray}
\begin{eqnarray}\label{hartree3}
V_{j,\alpha\sigma_{e}\tau_{Be}\alpha^{\prime}\sigma_{e}^{\prime}\tau_{Be}^{\prime}}^{HF}(\bm{r},\bm{r}^{\prime})
& = & \delta(\bm{r}-\bm{r}^{\prime})\sum_{j^{\prime}\beta\beta^{\prime}}\int
d\bm{s}\frac{Z_{j}Z_{j^{\prime}}}{|\bm{r}-\bm{s}|}\delta_{\sigma_{e}\sigma_{e}^{\prime}}\delta_{\tau_{Be}\tau_{Be}^{\prime}}\rho_{j^{\prime},\beta\sigma_{e}^{\prime\prime}\tau_{Be}^{\prime\prime}\beta\sigma_{e}^{\prime\prime}\tau_{Be}^{\prime\prime}}(\bm{s},\bm{s})\label{eq:HF_pot_Be7}\\
&  &
-\frac{Z_{j}^{2}}{|\bm{r}-\bm{r}^{\prime}|}\delta_{\alpha\alpha^{\prime}}\delta_{\sigma_{e}\sigma_{e}^{\prime}}\delta_{\tau_{Be}\tau_{Be}^{\prime}}\rho_{j,\alpha\sigma_{e}\tau_{Be}\alpha^{\prime}\sigma_{e}^{\prime}\tau_{Be}^{\prime}}(\bm{r},\bm{r}^{\prime})\nonumber
\end{eqnarray}
In Eqs. (\ref{hartree1}, \ref{hartree2}, \ref{hartree3}),
$\sigma_{e}\left(\frac{1}{2}\right)$ and $\tau_{Be}\left(\frac{3}{2}\right)$
are the electronic and nuclear spins, while $\alpha$ and $\beta$ represent
all the other quantum numbers. The index $j$ runs over all the fermionic particle types.
The self-consistent HF potential
is used in Eq. (\ref{gamma}), along with the external potential, $V=V_{HF}+V_{ext}$, to calculate the
(static exchange) electron-capture decay rate. \\
\begin{figure*}[t!!]
\includegraphics[width=1.1\linewidth]{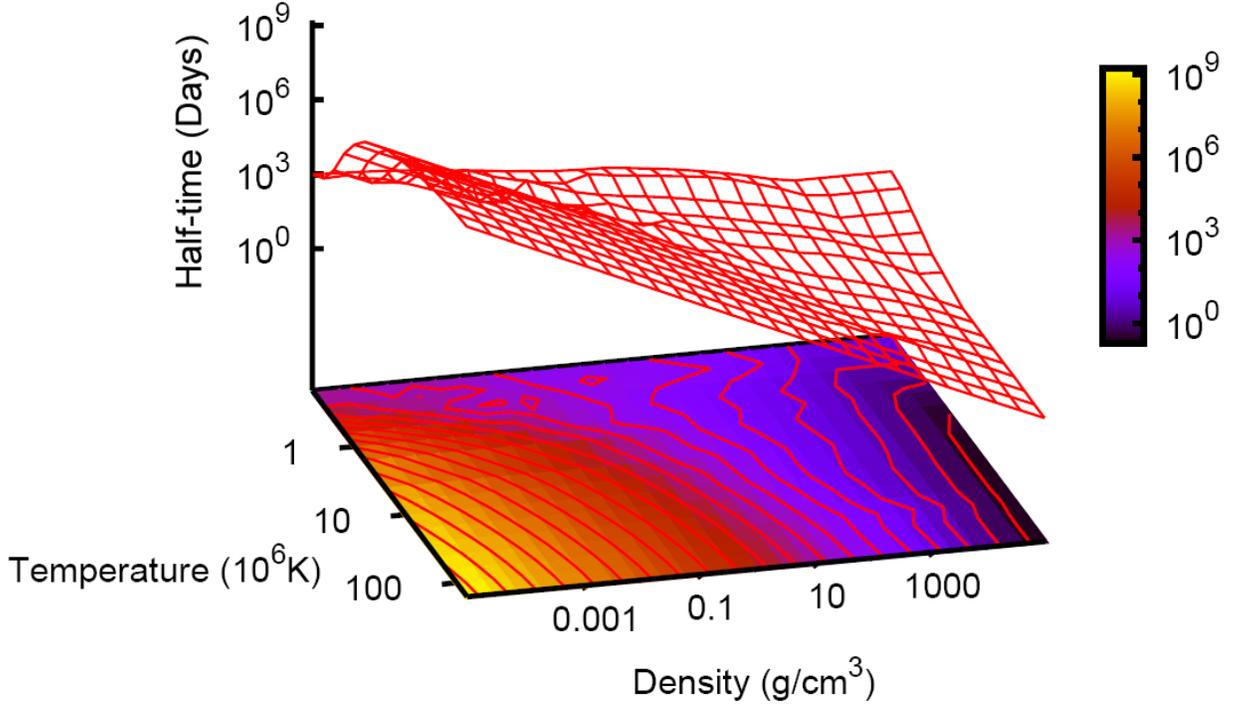}
\caption{The electron-capture half-life in days for $^7$Be as a function of $\rho$ and $T$.}
\label{decay}
\end{figure*}
\indent In order to be complete, the above treatment of the many-body interaction needs to include
the electron and proton continuum states, as the capture can occur from
the continuum orbitals and, at high $T$, the plasma is from partially to totally ionized.
To include the continuum states in the HF equations, we used the
theory of projected potentials, developed for the calculation of the electronic
emission spectra from solids \citep{compuscie,prb,taioli2010}.
Within this theory, while the Coulomb interaction among the particles (see
Eq. \ref{Hamiltonian}) is projected onto the Hilbert space
spanned by a basis set, giving rise to the discrete part of the spectrum, the kinetic terms are left unprojected.
In this way, as the eigenfunctions of the kinetic energy operator are plane waves
at a given energy, one can ``recover'' the energy continuum.
Furthermore, the sums over the discrete states $n$ in Eq. (\ref{hartree2})
should be thought of as integrals over the continuum. In our case a $cc-pVDZ$
Gaussian basis set (GBS), centered on the $^7$Be nucleus, has been optimized
for the calculation of the bound states. We underline that the addition of the
continuum states to the projected Coulomb interaction can have  significant
enhancement effects on the decay rate over a wide portion of the parameter
space spanned. This however does not include the solar case, where our
estimates are essentially indistinguishable from those reported by
\citet{Adelberger}.  We notice in any case that, when applicable, our changes
point to the opposite direction with respect to the suggestions by
\citet{Shaviv} for the Sun.\\
\indent Our two-body scattering framework allows us to go beyond even the HF
treatment of the screening; therefore, we estimated the importance of
correlations by computing $\rho_e(0)$ for an isolated $^7$Be atom using the
Full Configuration Interaction approach (FCI) for the previously used $cc-pVDZ$
GBS. We did not find any appreciable difference between the $\rho_e(0)$ estimates calculated by the HF and by the FCI methods, up to the third digit (see Table 3), thus justifying our mean-field approach, which neglects dynamical correlations.

\begin{table}[t]
\centering

{\large Table 3}

\begin{tabular}{c|c|c}
& Energy & $\rho_{e\uparrow}(0)$\tabularnewline
\hline
Hartree-Fock & $-14.5729914739$ & $17.68521$\tabularnewline
\hline
Full-CI & $-14.6604710493$ & $17.68060$\tabularnewline
\end{tabular}
%\label{tab3}
\caption{Energy of the isolated beryllium atom in atomic units and spin-up density at the nucleus obtained through the HF and CI calculations.}
\end{table}

The calculation of the decay rate, performed at different $T$ and $\rho$ values
using our approach is reported in Figure \ref{decay}.  We recall once again
that the above treatment of the many-body interaction goes beyond the DH
approximation previously used to estimate the electron-capture decay rate in
the Sun \citep{Iben,bahcall1962}. The field of applicability of the HF
screening effects can in fact be extended rigorously, to cover the whole range
of parameters found in the layers between the H-burning region and the envelope
base in evolved stars.

\section{An application to the Li production and destruction in evolved stars}

As an example of application of our new rate estimates to practical
nucleosynthesis problems in stars, we briefly illustrate here the
situation for rather low mass stars ($M  \lesssim 2 M_{\odot}$) of solar
metallicity, undergoing the RGB and AGB evolutionary phases in presence of
deep-mixing processes. They were recently discussed by \citet{Palmerini1,Palmerini2} within
the framework of reaction rates offered by \citet{Adelberger}.

We consider here only the impact on the above astrophysical scenario of our
"best choice" for the reaction rate (our new method), without analyzing the
individual behavior of all the four estimates presented, as we are preparing a
more thorough analysis on that, to appear in a forthcoming paper, also dealing
with various stellar masses and compositions.

\begin{figure}[hbt!]
\includegraphics[width=0.5\linewidth,angle=0]{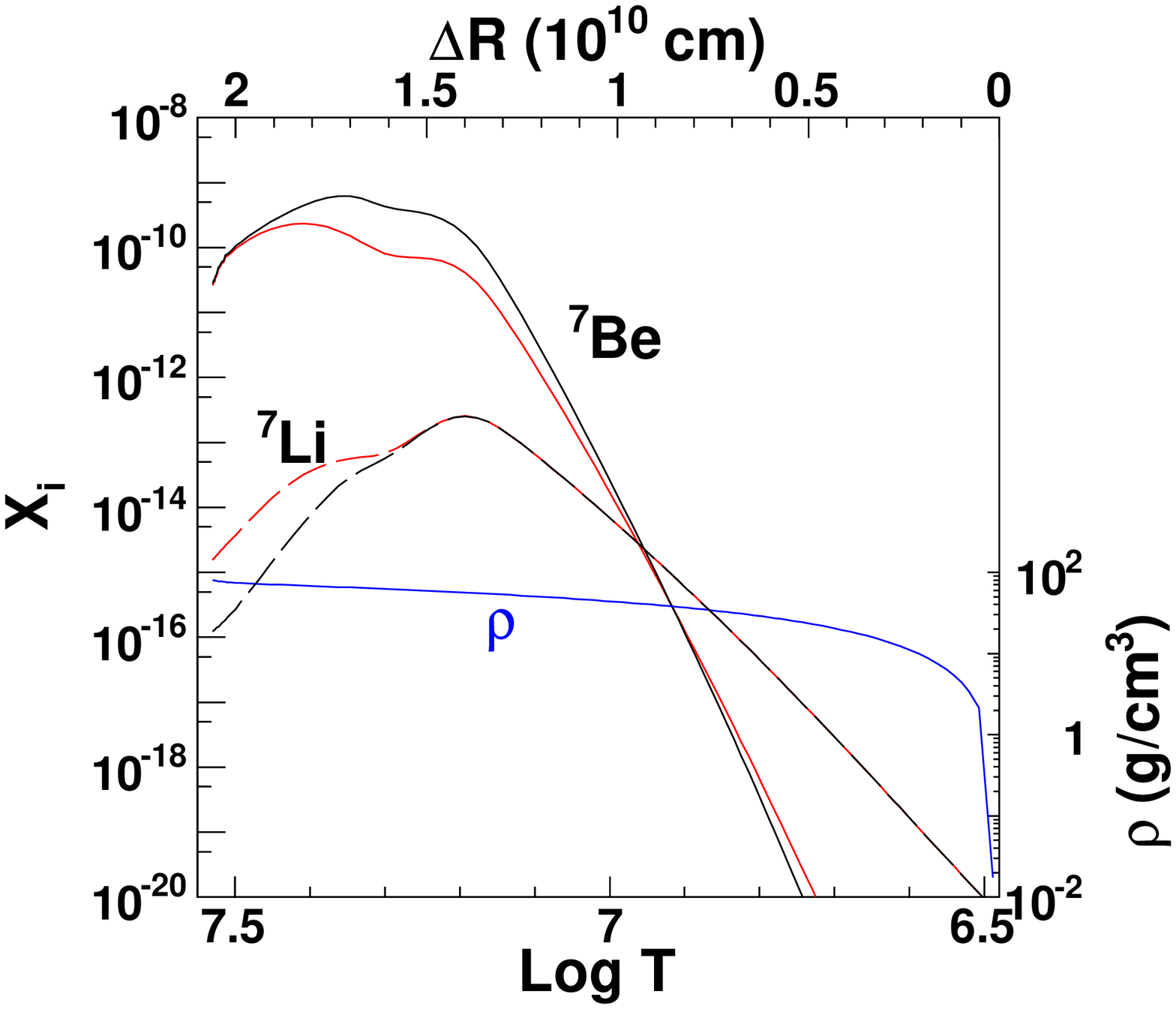}
\includegraphics[width=0.5\linewidth,angle=0]{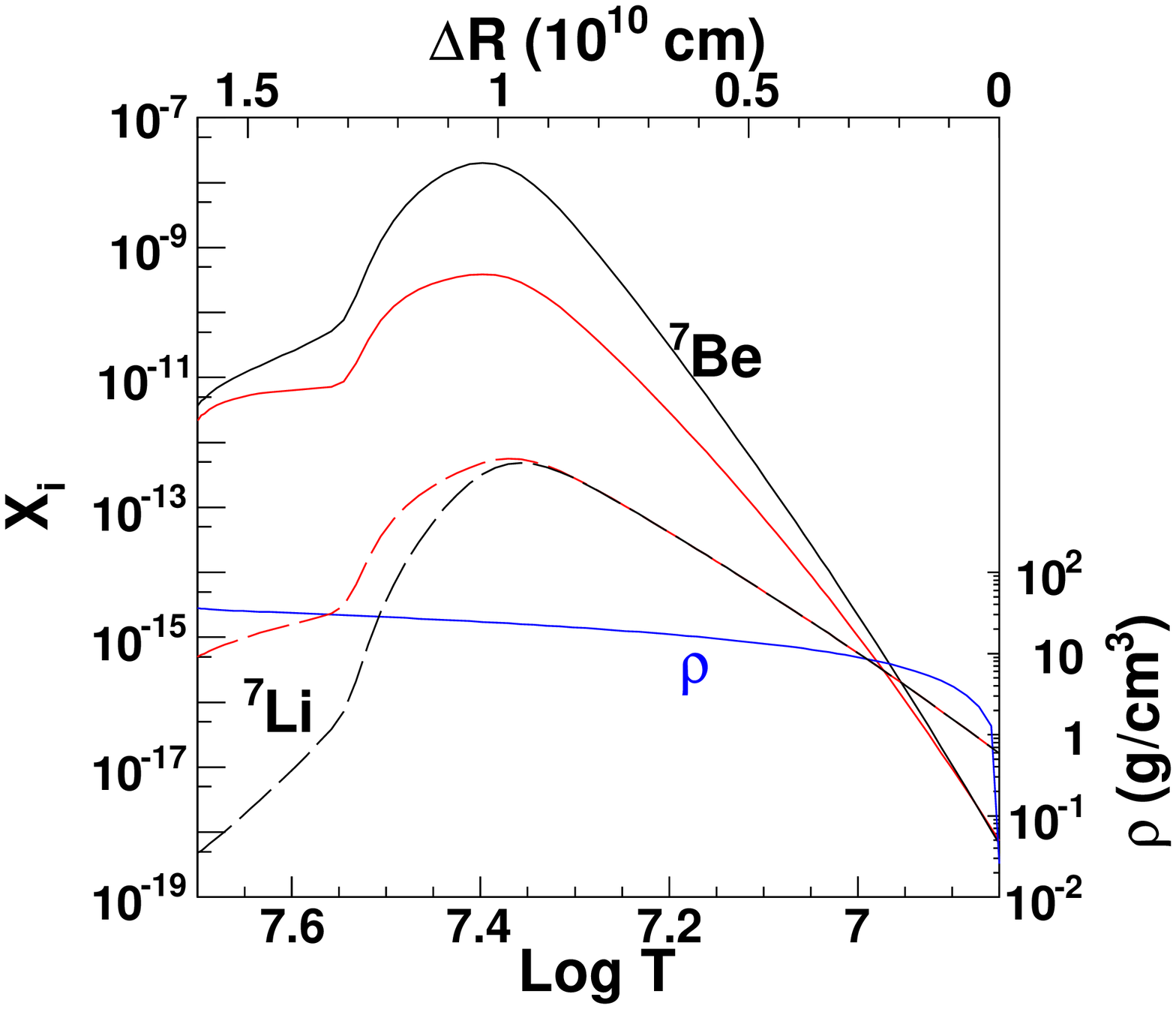}
\caption{A comparison between the equilibrium abundances of $^7$Be and $^7$Li
achieved in the layers above the H-burning shell, adopting our $^7$Be life-time
(red line) and  the one extrapolated by \citet{Adelberger} (black line),
in a 2 M$_{\odot}$ evolved star of solar metallicity. Panel a) refers to typical
RGB conditions, panel b) to AGB stages. The matter density is also shown (blue line),
and is referred to the scale on the right axis.}
      \label{shell}
\end{figure}

Figure \ref{shell} (a,b) shows the abundances of $^7$Be and of its daughter $^7$Li in the layers
below the convective envelope bottom and above the H-burning shell. They are plotted shortly
after the bump of the Luminosity Function on the RGB (panel a) and in between two thermal pulses,
during quiet H-burning, on the AGB (panel b). The stellar models are the same discussed in \citet{Palmerini2}.
The red lines illustrate the situation obtained with our new rate (the case previously labelled HF),
while the black line shows the previous findings, obtained with extrapolations
from \citet{Adelberger}, hence based on the DH approximation. The right-end
limit of the plot, at the position $\Delta r$ = 0, characterized by
temperatures below 10$^7$K, represents the base of the envelope; the left-end
limit is the region where the maximum energy is released from H-burning. As is
shown, the  matter
density of the layers considered never achieves the high values typical of the solar core: this is
the most critical reason why the DH approximation is inadequate for evolved
stars. In our more general approach both higher and lower values of the
electron density near the $^7$Be nucleus can be found, depending on the
conditions; however, over a major part of the region of interest, where the
density is not far from one tenth solar and the temperature remains high enough
($T \gtrsim 10^7$ K) the values of $\rho_e$(0) we find are higher than in the
DH model; the electron captures are therefore faster than in
\citet{Palmerini2}, and a lower equilibrium abundance of $^7$Be is established.

Notice that in our present example neither the $^3$He+$^4$He rate, nor the
$^7$Be+p or the $^7$Li+p rate have been modified, so that the variations shown
in Figure \ref{shell} are entirely due to the new approach adopted in computing
electron captures on $^7$Be. As discussed before, further effects are actually
to be expected on proton captures, as the changes in the electron density will
affect the screening of the Coulomb barrier for charged-particle interactions.

As discussed in the Section 1, it is known that interpreting the observed isotopic
abundances of light and intermediate elements in evolved stars does not require only the use
of the proper reaction rates, but also needs the assumption that deep mixing phenomena occur,
both destroying fragile nuclei  of the envelope (like Li), by exposing them to high temperatures,
and carrying to the surface products of H-burning nucleosynthesis.
These mixing mechanisms are generally parameterized and depend on two main parameters;
in the approach by \citet{Palmerini1,Palmerini2} these parameters are the mass circulation
rate (in units of $10^{-6}$ \ms/yr) induced by the transport processes ($\dot
M_6$) and the depth in the structure they achieve. This last is expressed in
terms of the (logarithmic) temperature difference between the deepest mixed
layers and those where the maximum energy is released in the H-shell ($\Delta =
\log T_H -\log T_P$). \citep[Recently, an analysis devised specifically for
fixing such extra-mixing parameters, adopting bright RGB stars as constraints,
was carried out by ][confirming the indications by Palmerini et al. on the need
of a rather shallow and slow mixing on the RGB]{abia}.

\begin{figure*}[hbt!]
\includegraphics[width=0.5\linewidth,angle=0]{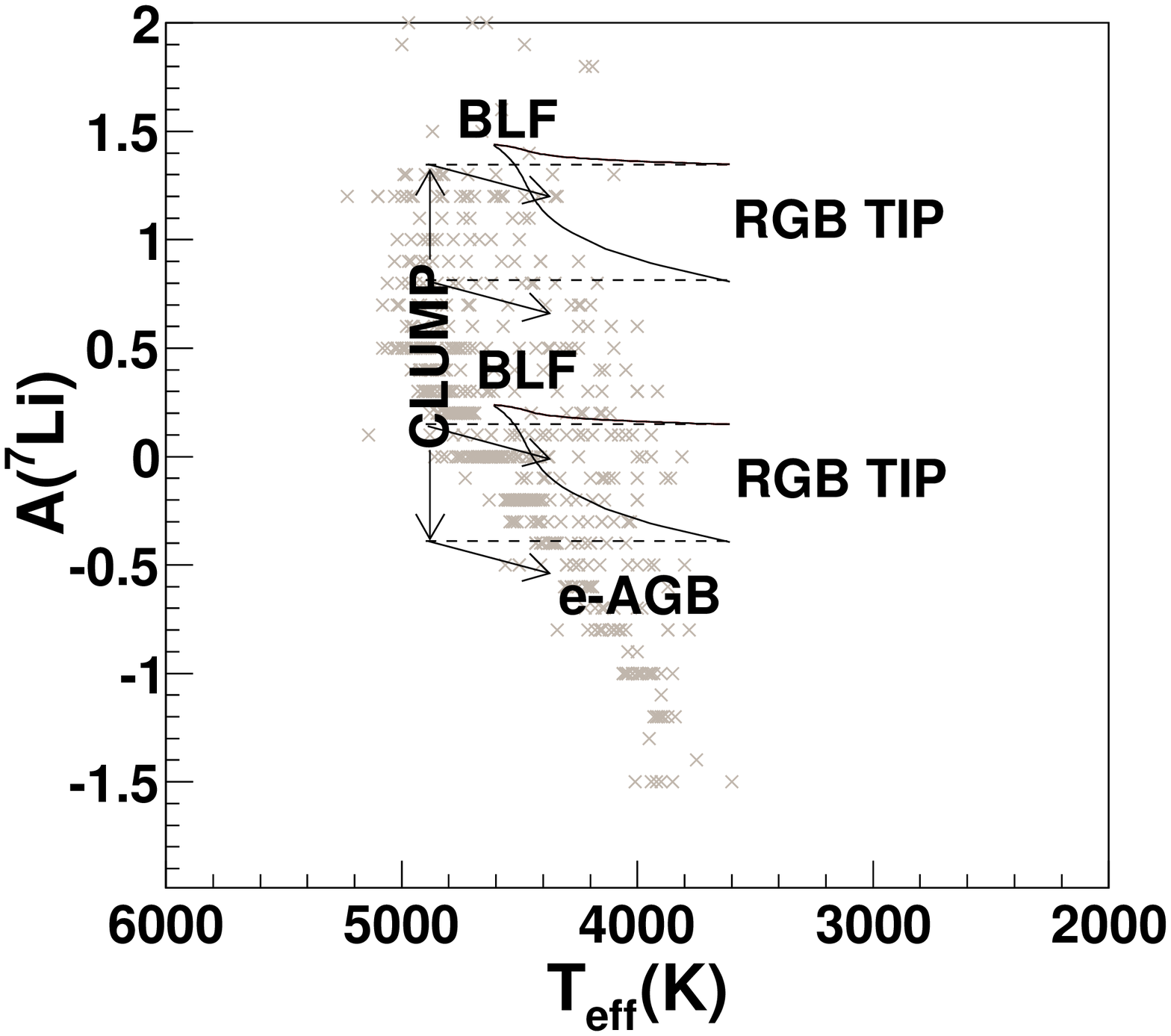}
\includegraphics[width=0.5\linewidth,angle=0]{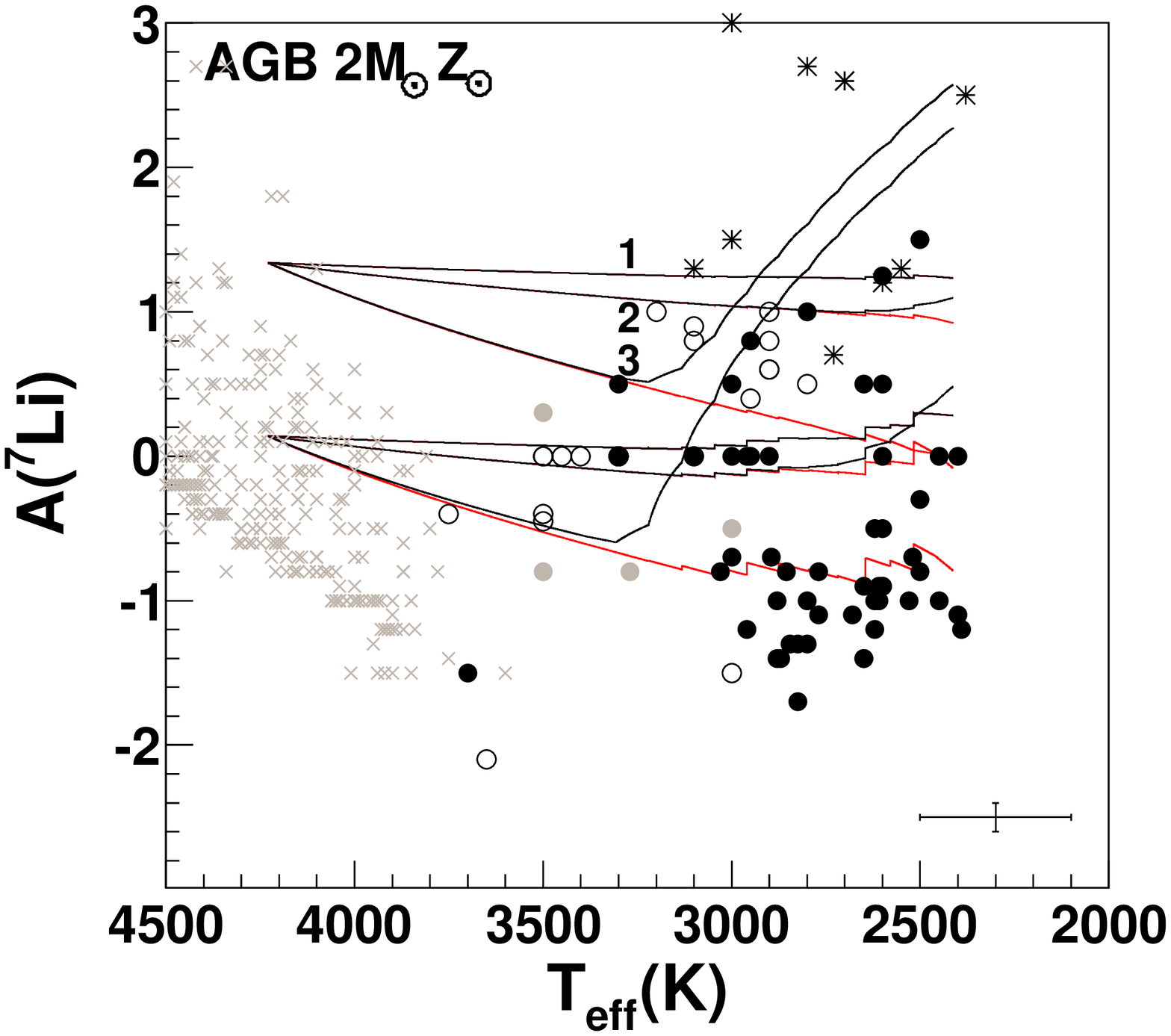}
\caption{Panel a) The Li destruction induced by extra-mixing on the RGB. Panel b) The
same effects for AGB  stages (see text for explanations, and \citet{Palmerini2} for the details of the observational points and their references)}
\label{mixing}
\end{figure*}

With reference to Panel a) in Figure \ref{mixing}, the points labeled BLF indicate the position in temperature of the Bump of the Luminosity Function, for two representative Li abundances, inside the wide spread allowed both by observations and models of the previous evolutionary stages (-0.3 $\lesssim A$(Li)$ \lesssim$ 1.5); crosses indicate observations. From each one
 of the two BLF conditions chosen, the two black curves show the Li destruction
 induced by adopting, for the extra-mixing parameters, $\Delta $ = 0.22
 and \mdot = 0.01 (upper curve) or 0.1 (lower curve). The effects of deep mixing
 are considered up to the moment when the RGB tip is reached. Subsequently, the
 temperature increases at constant values of $A$(Li) (dashed lines) up to the start of
 core He-burning (in the CLUMP region in the plot), after which Li would decrease again
 during the early-AGB stages
\citep[e-AGB in the plot; see also][for details]{Palmerini2}. The curves showing the
evolution of Li are completely insensitive to the adoption of the new or of the old
rate for $e^-$-captures. Panel b) of Figure \ref{mixing} shows instead that a different situation
emerges during the subsequent AGB evolution. Here we started the computations from three initial
values of the Li abundance, inside the observed spread left by RGB stages. Open and solid dots
in the Figure represent observations of M-MS-S and of C(N) stars, respectively. Asterisks show the
high Li abundances of CJ stars. The curves refer to stellar models with $\Delta$ = 0.22 and \mdot = 0.1, 0.3
1 (cases labelled 1, 2, 3, respectively).
The black curves are obtained using the $^7Be + e^-$ rate extrapolated by \citet{Adelberger},
while the red ones refer to our best choice from the present paper (HF). It is clear that
large changes emerge (in particular, with the new rate,  Li production becomes impossible).

\begin{figure*}[hbt!]
\includegraphics[width=0.5\linewidth,angle=0]{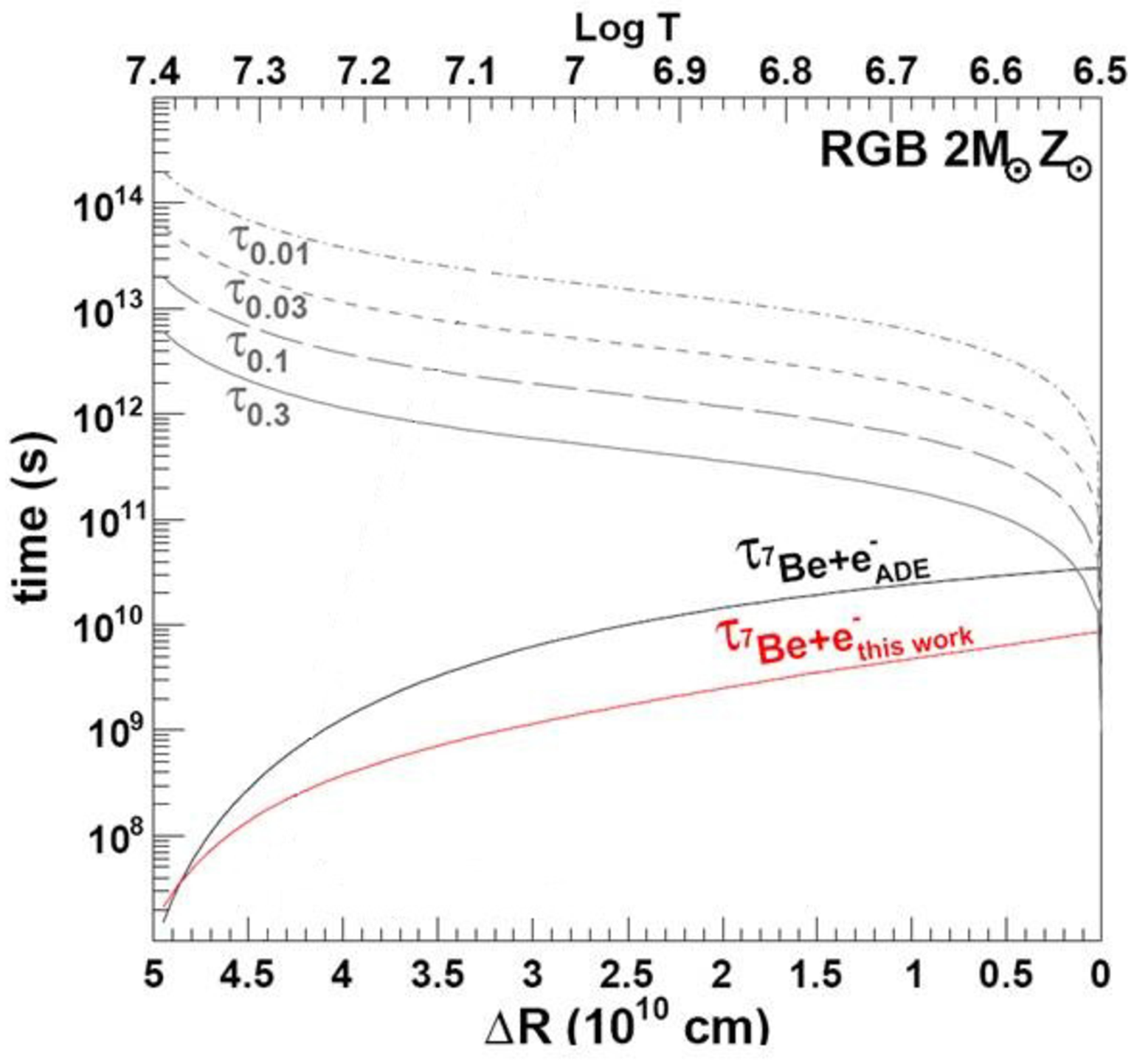}
\includegraphics[width=0.5\linewidth,angle=0]{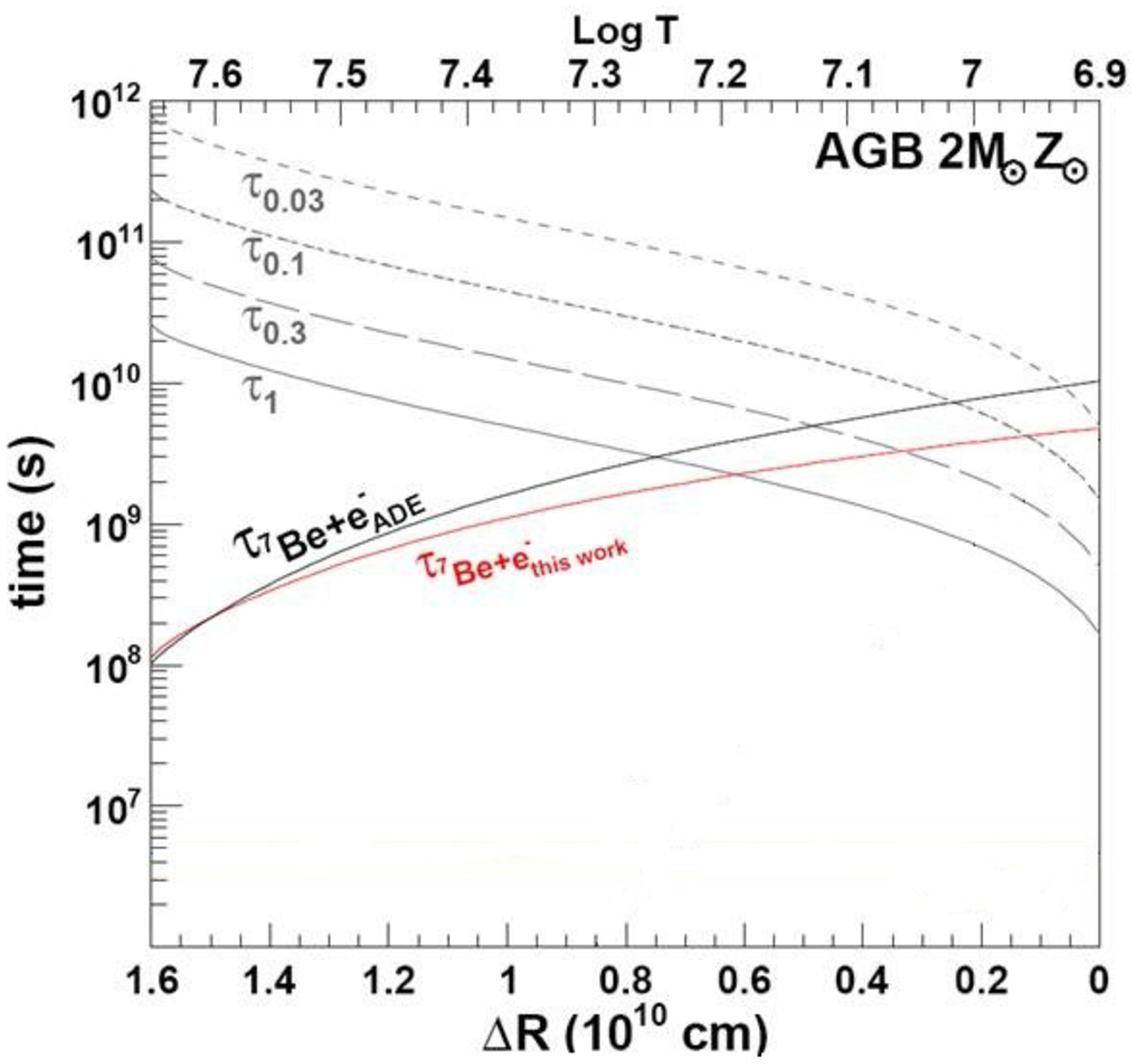}
\caption{A comparison between our life-time for $^7$Be, $\tau_{^7Be}$ (red line) and
the one extrapolated from \citet{Adelberger} (black line) with the time scales of
non-convective mixing at various rates (labelled by the mixing rate in units of $10^{-6}$
M$_{\odot}$/yr) in the radiative layers between the H-burning shell and the base of the
convective envelope, for a star with 2 M$_{\odot}$ and solar metallicity. Panel a) presents
a typical situation for the RGB phase, while panel b) is instead typical of AGB conditions.}
      \label{tau}
\end{figure*}

The reasons for the above dichotomy between RGB and AGB stages can be understood with
reference to the time scales for $^7$Be mixing and decay in the two cases. They are shown
in Figure \ref{tau}. As the figure shows, for the range of mixing rates required to explain
the other RGB chemical anomalies (panel a) both the new and old choice for the $^7$Be decay
provide a short enough lifetime for $^7$Be, to allow its decay to Li before being saved to
the envelope. In other words, conditions suitable for Be destruction are in any case met, both
with the older and with the newer choice of the rate. This explains why the new estimate for
the rate, which has evident effects in the stellar structure shown in panel a) of Fig \ref{shell},
does not modify the envelope abundances produced by the mixing process. The opposite occurs for
AGB stages (panel b). Here, the reduction in the lifetime of $^7$Be obtained with our new rate is
strong enough that even with relatively fast mixing processes the  destruction of Li in the
envelope is not compensated, and the highest observed Li abundances (typical of CJ stars) cannot
be reached. Essentially, with our rate Li behaves, on the AGB, in a way similar to what occurs on
the RGB: destruction always prevails.

We notice that, far from creating a new problem, this finding is now in agreement with the
indications provided by other chemical anomalies of the very peculiar CJ stars. Let's comment
on this in some more detail. In \citet{Palmerini2} it was surprisingly shown that, adopting the
extrapolation to AGB conditions of the $^7$Be lifetime from \citet{Adelberger}, the high
Li abundances observed in CJ-type carbon stars could be  explained in the framework of the
evolution of single stellar structures experiencing rather  fast extra-mixing
\citep[see in particular Fig. 10 in][]{Palmerini2}. This result was in itself quite strange, as
many peculiarities in such objects suggested instead, since a long time, that they do not follow
the ``normal'' evolutionary sequence for LMS. This is in particular demonstrated by their luminosity,
on average lower than for other C stars, by their lack of enrichment is s-process elements and by
the fact that a remarkable fraction of CJ stars have O-rich shells. Among the hypotheses presented
in the literature for explaining the anomalous CJ evolutionary path, binarity is probably the most
commonly-invoked scenario \citep{lam90,ai00}, especially if leading to coalescence into a single
peculiar object \citep{mcc}. In this case, obviously, the high Li abundances of the peculiar
descendants of such a stellar merging are not expected to be explained by single-star nucleosynthesis.
We actually predict that Li destruction, with our new rate, be even more facilitated than what emerges
from the first simple example of Figure \ref{mixing}. In fact a further effect is expected by an
increase in the electron screening, due to the now higher (on average) electron densities near the
nuclei, making proton captures on Li more effective.

In general, the Li-rich red giants (even outside the peculiar CJ class) need really
exceptionally fast transport rates to explain the net Li production. Their small number
says that these phenomena must be very rare, or concentrated in extremely short evolutionary
stages.

\section{Conclusions}
In this note we have presented a revision of the methods normally employed for
estimating electron captures and, more in general, the electron density near a nucleus in a
stellar plasma. For this scope we have devised an {\it ab-initio} technique, based on a
formalism that goes beyond the previously-adopted ones. In our method, we firstly reduced
the complicated many-body problem by a screened two-body scattering model,
by using an ``adiabatic'' factorization of the eigenfunctions, resembling the Born-Oppenheimer
(BO) approximation. We did that by fixing the reference frame into the $^7$Be nucleus and
then writing the different parts of the Hamiltonian in this frame. The frame itself was
considered as non-inertial, to account for the possibility that $^7$Be be
in motion in a rather complex way. In this coordinate transformation, the
non-inertiality translated into some technical complicacy, with the need of considering the
ensuing apparent forces. The mathematical details of this procedure were presented in an Appendix.
In our Hamiltonian we then had to include the electron kinetic and the electron-Be
potential energies, the proton kinetic and proton-Be potential energies, the Be
kinetic energy, the electron-proton, electron-electron, proton-proton
interactions. Furthermore, we also had to consider two-particle kinetic terms,
which identify the coordinates of the $j$-electron and $J$-proton relative to the $^7$Be coordinate.
In particular, the conditions under which the two terms coupling the Be and plasma
momenta must be considered carefully, as this is crucial for the application of
our mean-field treatment. We then looked for separable eigensolutions and
wrote the wavefunction factorization
in a way different from the usual formulation of the BO approximation. Indeed, at variance
with that approach, in our method there is neither the need that the electronic potential
energy surfaces be widely separated, nor that there is a mass scale difference among the
particles. By introducing the two-body framework and the relative coordinate
system, we achieved two main results. The first one is that in our approach
the $^7$Be nucleus is free to move around (thus overcoming the limitations of
the Born-Oppenheimer scheme); the second one is that the two-body electron-$^7$Be
density-matrix can be factorized as the product of two one-body density-matrices and thus
it is possible to introduce different schemes of approximations to the many-body interaction,
including the Hartree--Fock's one. Then, the screening effects due to all the fermions of
the environment, which modify the electron-$^7$Be scattering, were taken
into account, using standard many-body techniques.

By comparing our results with those obtained at various levels of simplification of the theory
we showed that, while the traditional DH approximation yields reliable results for the solar
core, where both high temperature and high density conditions exist, such simplified methods
do not always hold in stellar evolution. In particular, in the low density and
temperature environments characterizing partial H-burning above the H-shell in
Red Giants, where lithium undergoes important depletion, we showed that our
general method should be used and that extrapolations from existing
$e^-$-capture rates on $^7$Be in the Sun are not applicable.

As an example of the astrophysical consequences of our new method we recomputed, for a star
of $M$ = 2$M_{\odot}$ and solar metallicity, the evolution of the surface Li abundance in presence of deep mixing
mechanisms. While, for RGB phases, with the new technique we re-obtain previously known results, on the
AGB (where the density drops more remarkably) a stronger Li depletion was obtained and it was
shown that high Li abundances (as observed e.g. in CJ-type carbon-rich giants) cannot be the result of
extra-mixing occurring in single stars, but need more intricate evolutionary paths, as already
speculated in the past from other chemical peculiarities of these objects.

Apart from the limited example presented in this first contribution, our method yields,
in the regions characterized by temperatures of one to a few $10^7$K and
moderate densities (from 1 to 30-40 g/cm$^3$) higher values of the electron
density at the nucleus than the DH approach. These conditions are not relevant
for the solar core, but are important in red giant stars and in sub-envelope
layers of MS stars. This should have the effect of enhancing, in those
astrophysical environments, the electron screening, thus favoring thermonuclear
fusion. Such a possibility may be especially important for light nuclei (like
Li, or $^3$He), burning at relatively low temperature and that are related to
crucial cosmological problems, like the explanation of the gradual $^3$He
consumption in the Universe and of the huge Li destruction
below the surface of main sequence stars, including our Sun.

\acknowledgments

We thank Dr. G. Garberoglio (Bruno Kessler Foundation) for carefully reading the manuscript and
invaluable suggestions. ST gratefully acknowledges the Institute of Advanced Studies in Bologna
for the logistic support provided during his fellowship. SP acknowledges support from the
Spanish Grant AYA2011-22460

\appendix

\section{Transformation to relative coordinates}

The calculation of the electron-capture decay rate in a many-body system is a difficult task,
thus we propose to use an ``adiabatic'' approximation of the Hamiltonian,
which resembles the widely used Born-Oppenheimer
approach.

In order to derive the equation of motion for the fermions in the coordinate system relative to the
$^7Be$ nucleus, we will consider a model system composed by a single $^7Be$ atom surrounded
by $N_p$ protons and $N_e$ electrons. The Hamiltonian of such system in the laboratory frame can be written:
\begin{eqnarray}\label{Hamiltonian1}
H & = &
-\frac{1}{2M_{Be}}\nabla_{Be}^{2}+\sum_{j=1}^{N_{e}}\left(-\frac{1}{2m_{e}}\nabla_{e,j}^{2}\right)+\sum_{J=1}^{N_{b}}\left(-\frac{1}{2m_{p}}\nabla_{p,J}^{2}\right)-\sum_{j=1}^{N_{e}}\frac{Z_{Be}}{|\bm{R}_{Be}-\bm{r}_{e,j}|}\nonumber
\\ &  & +\sum_{J=1}^{N_{p}}\frac{Z_{Be}}{|\bm{R}_{Be}-\bm{R}_{p,J}|}
-\sum_{j=1}^{N_{e}}\sum_{J=1}^{N_{p}}\frac{1}{|\bm{r}_{e,j}-\bm{R}_{p,J}|}+\sum_{j=1}^{N_{e}}\sum_{k=j+1}^{N_{e}}\frac{1}{|\bm{r}_{e,j}-\bm{r}_{e,k}|}\nonumber
\\
&& +\sum_{J=1}^{N_{p}}\sum_{K=J+1}^{N_{p}}\frac{1}{|\bm{R}_{p,J}-\bm{R}_{p,K}|}
\end{eqnarray}
By performing the following transformation to the ($^7$Be-$e^-$ and $^7$Be-$p^+$) relative
coordinates
\begin{eqnarray}\label{transformation}
\bm{R}_{Be}^{\prime}&=&\bm{R}_{Be},\qquad\bm{r}_{e,j}^{\prime}=\bm{r}_{e,j}-\bm{R}_{Be},\qquad\bm{R}_{p,J}^{\prime}=\bm{R}_{p,J}-\bm{R}_{Be}\nonumber \\
\bm{\nabla}_{Be}&=&\bm{\nabla}_{Be}^{\prime}-\sum_{J=1}^{N_{p}}\bm{\nabla}_{p,J}^{\prime}-\sum_{j=1}^{N_{e}}\bm{\nabla}_{e,j}^{\prime},\qquad\bm{\nabla}_{e,j}^{\prime}=\bm{\nabla}_{e,j},\qquad \bm{\nabla}_{p,J}^{\prime}=\bm{\nabla}_{p,J}
\end{eqnarray}
one obtains:
\begin{eqnarray}\label{Hamfin}
&& H = \sum_{j=1}^{N_{e}}\left(-\frac{1}{2m_{e}}-\frac{1}{2M_{Be}}\right)\nabla_{e,j}^{\prime2}+\sum_{J=1}^{N_{p}}\left(-\frac{1}{2m_{p}}-\frac{1}{2M_{Be}}\right)\nabla_{p,J}^{\prime2}-\sum_{j=1}^{N_{e}}\frac{Z_{Be}}{|\bm{r}_{e,j}^{\prime}|}+\sum_{J=1}^{N_{p}}\frac{Z_{Be}}{|\bm{R}_{p,J}^{\prime}|}\nonumber \\
 && - \sum_{j=1}^{N_{e}}\sum_{J=1}^{N_{p}}\frac{1}{|\bm{r}_{e,j}^{\prime}-\bm{R}_{p,J}^{\prime}|}+\sum_{j=1}^{N_{e}}\sum_{k=j+1}^{N_{e}}\frac{1}{|\bm{r}_{e,j}^{\prime}-\bm{r}_{e,k}^{\prime}|}+\sum_{J=1}^{N_{p}}\sum_{K=J+1}^{N_{p}}\frac{1}{|\bm{R}_{p,J}^{\prime}-\bm{R}_{p,K}^{\prime}|} - \frac{1}{2M_{Be}}\nabla_{Be}^{\prime2} \nonumber \\
&& - \sum_{\substack{J,J^{\prime}=1 \\J \neq J^{\prime}}}
^{N_{p}}\left( \frac{1}{M_{Be}}\bm{\nabla}_{p,J}^{\prime}\cdot\bm{\nabla}_{p,J^{\prime}}^{\prime}\right) -\sum_{\substack{j,j^{\prime}=1 \\j \neq j^{\prime}}}^{N_{e}}\left(\frac{1}{M_{Be}}\bm{\nabla}_{e,j}^{\prime} \cdot \bm{\nabla}_{e,j^{\prime}}^{\prime}\right)-\frac{1}{M_{Be}}\sum_{j=1}^{N_{e}}\sum_{J=1}^{N_{p}}\bm{\nabla}_{p,J}^{\prime}\cdot\bm{\nabla}_{e,j}^{\prime} \nonumber \\
&& + \sum_{j=1}^{N_{e}}\left(\frac{1}{M_{Be}}\bm{\nabla}_{e,j}^{\prime}\cdot\bm{\nabla}_{Be}^{\prime}\right)+\sum_{J=1}^{N_{p}}\left(\frac{1}{M_{Be}}\bm{\nabla}_{p,J}^{\prime}\cdot\bm{\nabla}_{Be}^{\prime}\right)\nonumber \\
&&
\end{eqnarray}
Within this model the $^7$Be nucleus plays the role of the heavy nucleus in the BO approximation,
moving more slowly than the light particles (electrons and protons). \\
\indent Eq. (\ref{Hamfin}) can be simplified by neglecting all the inter-particle
coupling terms, which are divided by the Be mass ($M_{Be}$). However, the last two terms in Eq. (\ref{Hamfin}) are
special as they couple the Be nucleus and plasma momenta. Thus, we need to find the
conditions under which these two coupling terms can be neglected as this is crucial for the
application of our mean-field treatment.
To determine when this is possible, one can notice that the two-body density-matrix
of the system in relative coordinates at temperature $T$ is given by:
\begin{equation}\label{canonic}
\mathcal{Z=}\frac{1}{Z}\sum_{\alpha}\int\frac{d\bm{k}}{(2\pi)^{3}}e^{-\frac{E_{r,\alpha\bm{k}}}{k_{B}T}}  e^{-\frac{1}{2M_{\text{Be}}}\frac{k^{2}}{k_{B}T}} |\Phi_{\alpha,\bm{k}}><\Phi_{\alpha,\bm{k}}|\otimes|\chi_{\bm{k}}><\chi_{\bm{k}}|
\end{equation}
where $\alpha$ identifies the electronic
quantum numbers, $E_{r,\alpha\bm{k}}$ are the eigenvalues of the secular problem
for $\Phi$, and $Z$ is the canonical partition function.
The exponential term in Eq. (\ref{canonic}) kills the integral, unless:
\begin{equation}\label{kappa}
k\sim\sqrt{2M_{\text{Be}}k_{B}T}
\end{equation}
Using Eq. (\ref{kappa}), the two coupling terms in Eq. (\ref{Hamiltonian})
are of the order of $|\frac{1}{M_{Be}}\bm{k}\cdot\bm{\nabla}_{e,j}^{\prime}|\simeq 2\sqrt{k_{B}TK_em_e/M_{Be}}$
and $|\frac{1}{M_{Be}}\bm{k}\cdot\bm{\nabla}_{p,j}^{\prime}|\simeq 2\sqrt{k_{B}TK_pm_p/M_{Be}}$, where
$K_e$ and $K_p$ are the electron and proton kinetic energies, respectively.
Therefore, if the two conditions
\begin{equation}\label{conditions}
 4k_{B}Tm_e/M_{Be}\ll K_e, \quad 4k_{B}Tm_{p}/M_{Be}\ll K_{p}
 \end{equation}
hold, then
$\left(-\frac{i}{m_{e}}\bm{k}\cdot\bm{\nabla}_{e,j}^{\prime}\right)$
and
$\left(-\frac{i}{m_{p}}\bm{k}\cdot\bm{\nabla}_{p,j}^{\prime}\right)$
are negligible. These conditions are generally satisfied, due the presence
of the Be mass in the denominator and to the fact that
the scalar products $\left(-\frac{i}{m_{e}}\bm{k}\cdot\bm{\nabla}_{e,j}^{\prime}\right)$
and $\left(-\frac{i}{m_{p}}\bm{k}\cdot\bm{\nabla}_{p,j}^{\prime}\right)$
contain the cosine of the reciprocal direction of the two multiplying vectors,
which, on average, is very small.
Finally, the last three terms of the Hamiltonian (\ref{Hamfin}) can be safely
neglected as they contain a multiplying factor proportional to the inverse of
the Be mass.

\section{The Thomas--Fermi and Debye--H{\"u}ckel models}

The widely used DH model can be obtained via a two-step approximation starting from
the TF theory of the electron gas.
The TF model is a simplified HF theory, in which
the electron gas is treated within the local density approximation (LDA),
so that a large number of electrons is needed in a region where the potential is nearly constant.
In this theory the density is thus diagonal in the electron coordinates:
\begin{eqnarray}
\rho_{j,\alpha\alpha^{\prime}}(\bm{r},\bm{r}^{\prime}) & = & \rho_{j,\alpha\alpha^{\prime}}(\bm{r})\delta(\bm{r}-\bm{r}^{\prime})\\
\rho_{j,\alpha\alpha^{\prime}}(\bm{r}) & = & \int\frac{d\bm{p}}{(2\pi)^{3}}\frac{\psi_{j,\alpha\bm{p}}\psi_{j,\alpha^{\prime}\bm{p}}}{e^{\epsilon_{j,\bm{p}}}+1}
\end{eqnarray}
and can be obtained by the self-consistent solution of the TF equation:
\begin{equation}
-\frac{p_{j}^{2}}{2m_{j}}\psi_{j,\alpha\bm{p}}+V_{j,\bm{p}}^{ext}(\bm{r})\psi_{j,\alpha\bm{p}}-\mu_{j}\psi_{j,\alpha\bm{p}}+\sum_{\beta}V_{j,\alpha\bm{p}\beta\bm{p}}^{HF}\psi_{j,\beta\bm{p}}=\epsilon_{j,\bm{p}}\psi_{j,\alpha\bm{p}}
\end{equation}
where $V_{j,\bm{p}}^{ext}(\bm{r})$ is the electron-nucleus interaction, $\mu$ is the chemical potential,
$\alpha,\alpha^{\prime}$ identify the electron quantum numbers and
\begin{equation}
V_{j,\alpha\alpha^{\prime}}^{HF}(\bm{r},\bm{r}^{\prime}) = \delta(\bm{r}-\bm{r}^{\prime})\sum_{j^{\prime}\beta\beta^{\prime}}\int d\bm{s}g_{j\alpha\beta,j^{\prime}\alpha^{\prime}\beta^{\prime}}\left(\bm{r}-\bm{s}\right)\rho_{j^{\prime},\beta\beta}(\bm{s},\bm{s})-\sum_{\beta\beta^{\prime}}g_{j\alpha\beta,,j\beta^{\prime}\alpha^{\prime}}(\bm{r}-\bm{r}^{\prime})\rho_{j,\beta\beta^{\prime}}(\bm{r},\bm{r}^{\prime})
\end{equation}.
Here $g$ is the bare Coulomb potential.\\
\indent In order to obtain the DH approximation, in the first step
one substitutes the quantum-mechanical Fermi-Dirac statistics with the classical Boltzmann
distribution, while, in a second step, the high-temperature (weak coupling)
limit is obtained by Taylor expanding at the first order the exponential distribution:
\begin{equation}
\rho_{j}(\bm{r})=\int\frac{d\bm{p}}{(2\pi)^{3}}e^{-\beta\left[\frac{p_{j}^{2}}{2m_{j}}+q_{j}\Phi(\bm{r})-\mu_{j}\right]}
\end{equation}
In this way one obtains the DH equation for a neutral plasma as follows:
\begin{equation}\label{Debye}
\nabla^2 \Phi = \lambda_D \Phi
\end{equation}
where
\[
\Phi(\bm{r})=\sum_{j^{\prime}}\int d\bm{r}^{\prime}\frac{q_{j^{\prime}}\rho_{j}(\bm{r})}{|\bm{r}-\bm{r}^{\prime}|}
\]
In Eq. (\ref{Debye}), the DH length is defined by
$\lambda_D=(\epsilon_r\epsilon_0k_BT/\sum_{j=1}^N\rho_j^0q_j^2)^{\frac{1}{2}}$,
where $k_B$ is the Boltzmann constant, $\rho_j^0$ is the mean charge density of the species $j$, $\epsilon_r$ and $\epsilon_0$
are the relative and vacuum dielectric constants. In this framework, $\lambda_D$ sets the characteristic length scale for the variation
of the potential and of the charge concentration.

{\bf Acknowledgements.} We are grateful to an unknown referee for very pertinent and helpful comments, which we believe were crucial for improving the quality of the presentation.

\begin{landscape}
\footnotesize
\tightenlines
\centerline{\large Table 2}
\begin{tabular}{|c|c|c|c|c|c|c|c|}
\hline
$\rho$ ($g/cm^{3}$) & $T$ ($10^{6}K$) & $\lambda_{Debye}$ $a.u.$ & $\lambda_{De\, Broglie}$ $(e-p)$ & $\rho_{HF}(0)$ $a.u.$ & $\rho_{TF}(0)$ & $\rho_{B}(0)$ & $\rho_{DH}(0)$\tabularnewline
\hline
\hline
$1000.$ & \multirow{7}{*}{$1.$} & $0.038$ & \multirow{7}{*}{$1.409$ - $0.0329$} & $71.87\div71.97$ & $68.99\div69.11$ & $42.61\div42.74$ & $47.46\div47.55$\tabularnewline
\cline{1-1} \cline{3-3} \cline{5-8}
$100.$ &  & $0.119$ &  & $33.52\div33.53$ & $29.53\div29.55$ & $4.027\div4.031$ & $19.13\div19.14$\tabularnewline
\cline{1-1} \cline{3-3} \cline{5-8}
$10.$ &  & $0.377$ &  & $17.37\div17.37$ & $13.83\div13.83$ & $0.945\div0.945$ & $13.33\div13.33$\tabularnewline
\cline{1-1} \cline{3-3} \cline{5-8}
$1.$ &  & $1.193$ &  & $7.839\div7.837$ & $5.708\div5.707$ & $0.184\div0.184$ & $8.151\div8.149$\tabularnewline
\cline{1-1} \cline{3-3} \cline{5-8}
$0.1$ &  & $3.771$ &  & $1.940\div1.940$ & $1.415\div1.415$ & $0.044\div0.044$ & $2.059\div2.058$\tabularnewline
\cline{1-1} \cline{3-3} \cline{5-8}
$0.01$ &  & $11.93$ &  & $0.278\div0.278$ & $0.220\div0.220$ & $0.0075\div0.0075$ & $0.279\div0.279$\tabularnewline
\cline{1-1} \cline{3-3} \cline{5-8}
$0.001$ &  & $37.71$ &  & $0.0308\div0.0308$ & $0.0264\div0.264$ & $0.0012\div0.0012$ & $0.0303\div0.303$\tabularnewline
\hline
$1000.$ & \multirow{7}{*}{$10.$} & $0.119$ & \multirow{7}{*}{$0.445$ - $0.0103$} & $122.43\div122.89$ & $116.21\div116.68$ & $51.77\div52.05$ & $108.56\div109.01$\tabularnewline
\cline{1-1} \cline{3-3} \cline{5-8}
$100.$ &  & $0.377$ &  & $20.23\div20.27$ & $19.53\div19.57$ & $10.36\div10.39$ & $19.54\div19.58$\tabularnewline
\cline{1-1} \cline{3-3} \cline{5-8}
$10.$ &  & $1.193$ &  & $2.578\div2.581$ & $2.554\div2.558$ & $2.515\div2.519$ & $2.570\div2.573$\tabularnewline
\cline{1-1} \cline{3-3} \cline{5-8}
$1.$ &  & $3.771$ &  & $0.274\div0.275$ & $0.274\div0.275$ & $0.274\div0.274$ & $0.274\div0.275$\tabularnewline
\cline{1-1} \cline{3-3} \cline{5-8}
$0.1$ &  & $11.93$ &  & $0.0281\div0.0282$ & $0.0281\div0.0282$ & $0.0281\div0.0282$ & $0.0281\div0.0281$\tabularnewline
\cline{1-1} \cline{3-3} \cline{5-8}
$0.01$ &  & $37.71$ &  & $\left(2.84\div2.84\right)\cdot10^{-3}$ & $\left(2.84\div2.84\right)\cdot10^{-3}$ & $\left(2.84\div2.84\right)\cdot10^{-3}$ & $\left(2.83\div2.83\right)\cdot10^{-3}$\tabularnewline
\cline{1-1} \cline{3-3} \cline{5-8}
$0.001$ &  & $119.3$ &  & $\left(2.84\div2.84\right)\cdot10^{-4}$ & $\left(2.84\div2.84\right)\cdot10^{-4}$ & $\left(2.84\div2.84\right)\cdot10^{-4}$ & $\left(2.84\div2.84\right)\cdot10^{-4}$\tabularnewline
\hline
$1000.$ & \multirow{7}{*}{$100.$} & $0.377$ & \multirow{7}{*}{$0.141$ - $0.0033$} & $78.31\div80.39$ & $78.24\div\div80.32$ & $76.57\div78.64$ & $78.22\div80.30$\tabularnewline
\cline{1-1} \cline{3-3} \cline{5-8}
$100.$ &  & $1.193$ &  & $9.051\div9.289$ & $9.051\div9.288$ & $9.031\div9.268$ & $9.051\div9.288$\tabularnewline
\cline{1-1} \cline{3-3} \cline{5-8}
$10.$ &  & $3.771$ &  & $0.773\div0.787$ & $0.773\div0.787$ & $0.773\div0.787$ & $0.773\div0.787$\tabularnewline
\cline{1-1} \cline{3-3} \cline{5-8}
$1.$ &  & $11.93$ &  & $0.0775\div0.0789$ & $0.0775\div0.0789$ & $0.0775\div0.0789$ & $0.0775\div0.0789$\tabularnewline
\cline{1-1} \cline{3-3} \cline{5-8}
$0.1$ &  & $37.71$ &  & $\left(7.75\div7.90\right)\cdot10^{-3}$ & $\left(7.75\div7.90\right)\cdot10^{-3}$ & $\left(7.75\div7.90\right)\cdot10^{-3}$ & $\left(7.75\div7.90\right)\cdot10^{-3}$\tabularnewline
\cline{1-1} \cline{3-3} \cline{5-8}
$0.01$ &  & $119.3$ &  & $\left(7.75\div7.90\right)\cdot10^{-4}$ & $\left(7.75\div7.90\right)\cdot10^{-4}$ & $\left(7.75\div7.90\right)\cdot10^{-4}$ & $\left(7.75\div7.90\right)\cdot10^{-4}$\tabularnewline
\cline{1-1} \cline{3-3} \cline{5-8}
$0.001$ &  & $377.1$ &  & $\left(7.75\div7.90\right)\cdot10^{-5}$ & $\left(7.75\div7.90\right)\cdot10^{-5}$ & $\left(7.75\div7.90\right)\cdot10^{-5}$ & $\left(7.75\div7.90\right)\cdot10^{-5}$\tabularnewline
\hline
\end{tabular}
\centering{Values of the electron density $\rho_e(0)$ (in atomic units) at the Be nucleus, for different theoretical approaches (see text for explanations)}
\end{landscape}

\end{document}